\documentclass[pdflatex, sn-apa]{sn-jnl}
\usepackage{siunitx}
\jyear{2022}%

\theoremstyle{thmstyleone}%
\theoremstyle{thmstyletwo}%

\theoremstyle{thmstylethree}%

\begin{document}

\title[Neuropunk Revolution. Hacking Cognitive Systems towards Cyborgs 3.0]{Neuropunk Revolution. Hacking Cognitive Systems towards Cyborgs 3.0}


\author[1]{\fnm{Max} \sur{Talanov}}\email{max.talanov@gmail.com}

\author[2]{\fnm{Jordi} \sur{Vallverd\'u}}\email{jordi.vallverdu@uab.cat}

\author*[1]{\fnm{Andrew} \sur{Adamatzky}}\email{andrew.adamatzky@uwe.ac.uk}

\author[3]{\fnm{Alexander} \sur{Toschev}}\email{alexander.toschev@gmail.com}

\author[3]{\fnm{Alina} \sur{Suleimanova}}\email{sulemanovaaa@gmail.com}

\author[3]{\fnm{Alexey} \sur{Leukhin}}\email{alexey.panzer@gmail.com}

\author[3]{\fnm{Ann} \sur{Posdeeva}}\email{keiko.persik@gmail.com}

\author[3]{\fnm{Yulia} \sur{Mikhailova}}\email{mihaylova.yuliyaa@gmail.com}

\author[3]{\fnm{Alice} \sur{Rodionova}}\email{alice.palada@gmail.com}

\author[4]{\fnm{Alexey} \sur{Mikhaylov}}\email{mian@nifti.unn.ru}

\author[5]{\fnm{Alexander} \sur{Serb}}\email{A.Serb@soton.ac.uk}

\author[4]{\fnm{Sergey} \sur{Shchanikov}}\email{seach@inbox.ru}

\author[4]{\fnm{Svetlana} \sur{Gerasimova}}\email{gerasimova@neuro.nnov.ru}

\author[6]{\fnm{Mohammad Mahdi} \sur{Dehshibi}}\email{mdehshibi@uoc.edu}

\author[7]{\fnm{Alexander} \sur{Hramov}}\email{a.hramov@innopolis.ru}

\author[4]{\fnm{Victor} \sur{Kazantsev}}\email{kazantsev@neuro.nnov.ru}

\author[8]{\fnm{Tatyana} \sur{Tsoy}}\email{tt@it.kfu.ru}

\author[8]{\fnm{Evgeni} \sur{Magid}}\email{dr.e.magid@ieee.org}

\author[9]{\fnm{Igor} \sur{Lavrov}}\email{igor.lavrov@mayo.edu}

\author[10]{\fnm{Victor} \sur{Erokhin}}\email{victor.erokhin@imem.cnr.it}

\author[11]{\fnm{Kevin} \sur{Warwick}}\email{aa9839@coventry.ac.uk}

\affil*[1]{\orgdiv{Unconventional Computing Lab}, \orgname{University of the West of England}, \orgaddress{\street{Frenchay Campus, Coldharbour Lane}, \city{Bristol},  \country{UK}}}

\affil[2]{\orgdiv{de Filosofia}, \orgname{ICREA Acadèmia-UAB}, 
\orgaddress{
\city{Barcelona}, \country{Spain}}
}

\affil[3]{\orgname{B-Rain Labs LLC}, 
\orgaddress{\street{Profsouznaya st. 40-42}, \city{Kazan}, \country{Russia}}}

\affil[4]{\orgname{Lobachevsky University}, 
\orgaddress{\street{23 Gagarin prospect}, \city{Nizhny Novgorod}, \country{Russia}}}

\affil[5]{\orgname{University of Southampton}, \orgaddress{\street{Highfield campus}, \city{Southampton}, \country{UK}}}

\affil[6]{\orgname{Universitat Oberta de Catalunya}, \orgaddress{\street{Rambla del Poblenou, 156}, \city{Barcelona}, \country{Spain}}}

\affil[7]{\orgname{Innopolis University}, \orgaddress{\street{Universitetskaja str.}, \city{Innopolis}, \country{Russia}}}

\affil[8]{\orgname{Kazan Federal University}, \orgaddress{\street{Kremlevskaya 35}, \city{Kazan}, \country{Russia}}}

\affil[9]{\orgname{Mayo Clinic}, \orgaddress{\street{215 Highland CT SW Rochester}, \city{Rochester}, \country{USA}}}

\affil[10]{\orgname{IMEM- CNR}, \orgaddress{\street{Parco Area delle Scienze 37A}, \city{Parma}, \country{Italy}}}

\affil[11]{\orgname{Coventry University}, \orgaddress{\street{Priory Street}, \city{Coventry}, \country{UK}}}


\abstract{This work is dedicated to the review and perspective of the new direction that we call ``Neuropunk revolution'' resembling the cultural phenomenon of cyberpunk. 
This new phenomenon has its foundations in advances in neuromorphic technologies including memristive and bio-plausible simulations, BCI, and neurointerfaces as well as  unconventional approaches to AI and computing in general. We present the review of the current state-of-the-art and our vision of near future development of scientific approaches and future technologies.
We call the ``Neuropunk revolution'' the set of trends that in our view provide the necessary background for the new generation of approaches technologies to integrate the cybernetic objects with biological tissues in close loop system as well as robotic systems inspired by the biological processes again integrated with biological objects.
We see bio-plausible simulations implemented by digital computers or spiking networks memristive hardware as promising bridge or middleware between digital and [neuro]biological domains.}




\maketitle

\section{Introduction}\label{sec:intro}

In this paper we suggest a new engineering and conceptual way to hack human nervous thus cognitive systems, but before we explain the details of such project we should revise briefly the precedents and fundamental steps that paved the way to such historical moment. 

Creating languages as symbolic information conveyors was one of the fundamental moments in the history of human evolution. The next huge step in the history of humanity was the moment in which, thanks to such tools and their related possibilities, our ancestors were able to hack other cognitive systems (rhetorically) and, at the end, even enhance their own cognitive performance (thanks to symbolic tools like writing systems, mathematics, formal reasoning or logics). 
Firstly, humans manipulated living systems at macro (breeding selection, mammals domestication, plants grafting, etc) although operated with micro systems, without real knowledge of them, like yeast and other fungi, and secondly, at a  micro level (synthetic biology), when were able to create living systems \textit{ex novo} (minimal cells) using several techniques. 
Now, humans explore the integration of biological systems and technological devices though a variate range of technologies, like neuroimplants, prosthetic devices, chip implants 
(Kevin Warwick's firstly implanted the silicon chip in 1998, later he extended nervous system with a third robotic arm connected to his arm via neural interface~\citep{vogel2002part}), 
or brain-machine magnetic connections. 
Thanks to the current advances in the understanding of brain performance, as well as to the new technologies that allow us to modify neurochemical communication, a new revolution is in front of us: we've suggested to label it as ``the neuropunk revolution''. 
We don't want to discuss the huge meanings and conceptual load of the ``revolution'' concept \citep{sep-scientific-revolutions} by  this term  we mean the temporal context in which a significative change or advance is achieved in some scientific field, like the events of the Copernican revolution (1573), Bacon's creation of scientific method (1620), the Newtonian universal laws of gravity (1687), or the neuronal discoveries of Santiago Ram\'on y Cajal (Nobel prize in 1906). 
We suppose that the complex of phenomena described further in the article is supported by what has been called the Fourth Industrial Revolution \citep{schwab2017fourth}, as  this fourth era supports technologies that combine hardware, software, and biology (cyber-physical systems). 
The aim of this paper to describe the way of integrate bidirectionally machines and living systems, with special interest in mammals, and above all, human beings. 
The mechanism that will allow it is based on the neuronal performance, distributed all through the brain and the different sections of the nervous system. Finally, the inclusion of the ``punk'' term is related to some conceptual aspects of our research:  in the same way that punk movement defended  anti-establishment views, our research pushes against the historically and culturally defended boundaries of the so-called ``natural body''; and it is a ``neuropunk'' revolution because it is possible thanks  to the advances mediated by the knowledge about the Central Nervous System, and the understanding of the brain architecture and morphological functioning. At he same time, some echoes of the classic science fiction term ``cyberpunk'' remain here, but instead of considering our contribution as of part of such a dystopian futuristic setting, we only see the beneficial aspects of those enhancements. Therefore, and because of all the previously arguments, defend the conceptual value of the term ``cyberpunk revolution'', as a innovative, technology-mediated, and positive step into the advance of both human cognition and sensory-motor operational skills.

In following sections we define the nature of such technological mixtures (2), with details about the approach of this paper (3),  with special analysis of the design of bio-plausible simulations through software and hardware (spiking neurons and NNs, memristive synapses, FPGA) (4). In section (5) an exploration of machine to biological system interfaces will be explored, with special emphasis on (6) brain-computer interfaces. 
After such technical details we will debate (6) the revolution in this particular approach.
In section 7 we will analyze the value of considering unconventional AI as a path for such revolution and its possible implementation into robots (8). 
The last 9th section is dedicated to Challenges and limitations in  development of neuroimplants and BCI techniques.
The perspective review will conclude with the description of strong remarks and achieved advances.

\section{Biotechnical hybrids}\label{sec:chimeras}

The attempt to access a neural network of living systems is not a new one. 
Although Ancient Greeks discovered indirectly the existence of the Human nervous System \citep{panegyres2016ancient}, 
only the studies of Galvani 
and his fruitful controversy with Volta about the difference between animal and physical electricity \citep{piccolino1997luigi}, 
when a nervous system, that of a frog (basically, the contact of two different metals with the leg muscles of a skinned frog resulted in the generation of an electric current that caused the leg to twitch), 
was hacked through the application of electricity . 
The phenomenon was called ``animal electricity'' paved the way for the defense of mechanism, as a new cosmovision about life. 
Beyond historical details, electrophysiology was then born, and also, the way to identify ways to connect machines and bodies.
This amazing possibility, to move and ``resurrect'' dead animals, created a huge impact in European societies, as the book of \textit{Frankenstein; or, The Modern Prometheus.} of Mary Shelley (1818)  captured from a literary point of view. Such attempts to modify life entering into the bodies experienced a huge step forward when biochemistry innovations allowed to increase the detail of living machinery. 
The 20th century biochemical revolution allowed the rise of molecular biology, and at the end of the century genetic engineering, first, and synthetic biology, later, reduced the size of the modifiable aspects of living systems allowing the introduction of an engineering perspective into live design. Nevertheless, such narrow approach to life, beyond a systemic perspective of living systems, produced some fundamental problems when such systems where hacked \citep{gustafsson2016best}. 
The question regarding the modifiable nature of living systems was not the problem, as life is immersed in a natural creation of chimeras \citep{margulis2011chimeras},  but how the changes were integrated correctly into the main biological system. 
Our research attempts to explore the more sophisticated and modular cognitive system: human brain.
We face one fundamental problem: how can we create smooth information channels between our machines and our brains? 
Since 1973, when the term ``brain-computer interface'' was introduced \citep{vidal1973toward}, several methods have been created to connect brains to engineering systems (computers, prosthesis,~etc). 
Thanks to the fact that the human mind is highly adaptive to new ways of performing cognitive tasks, this skill allows remapping the embodiment and shows how the cognitive process is an enactive evolutionary mechanism. 
By the term ``enactivism'' we mean the current version of recently developed embodied cognition approaches. 
This view of cognitive processes are deeply entangled in action, instead of internal and representational ways of representing the world. 
Therefore, the process creating meaning or significance is fully embodied \citep{de2010enaction}.

In order to hack natural pathways and mechanisms of human cognitive system, our model of the closed loop system, to be discussed in detail in next sections, allows a new step into biohacking, much more deep and with more precision than any other previous system. 
Current biohacking attempts \citep{yetisen2018biohacking} are the example of non-systematic approach to hack the nervous system, and mainly based on simple uses of biological devices. 
This ``Do-It-Yourself Biology'' or ``garage biology'' \citep{biohacker2015biohacker} is of course not related to another of the fundamental meanings of this concept: the biological experimentation (as by gene editing or the use of drugs or implants) done to improve the qualities or capabilities of living organisms especially by individuals and groups working outside a traditional medical or scientific research environment \citep{warwick20204}. 
In order to avoid such possible confusion about the use of biohacking technologies, our research suggests the use of a different concept: ``technohacking''. 
Despite of existing as an artistic form of some collectives, this term has not still used in a scientific context. 
The term ``technohacking'' fits better to our approach: technological systems used to hack biological devices, from an engineering perspective, and combining several models and possible techniques. On the other hand current medical brain implants or prosthetic mechanisms lack of a generic interface to design upgraded humans, while present some ethical problems, like identity persistence. 
The selection of this term has been accurate in relation to our point of view about the implications of our model: it is not just a mere upgrade of the possible technological interventions done between humans and machines, but a direct modification, whose action is more relation to a smooth hacking than  a modification or a mere intervention.

Our model allows to identify and adjust clearly the functional aspects of the whole system of the human-machine integration. The closed loop system allows the full explainability of the functional properties of the resulting system. Such revolution allows to redefine next steps towards a neuropunk revolution based on a controlled transhumanism, as a second reliable wave towards the achievement of augmented humans or, as we call it, ``Project Cyborg 3.0''. Up to now, partial attempts to upgrade human bodies (at a very high medical cost) had been carried on, but our model makes the full process not only much more simple, but also effective, and controllable. Recent projects, like Neuralink are much more invasive, and costly, as compared to our, presented later approach~\citep{pisarchik2011optical}.

Before we dive in the details we should look at previous leading researches in the field. As part of his innovative research project ``Cyborg 2.0.'', on the 14th of March 2002 a one hundred electrode array was surgically implanted into the median nerve fibres of the left arm of Professor Kevin Warwick, for example he controlled a third robotic arm using it.
20 years later, we suggest a new non-invasive way to deal with human technological enhancement. 
According to his own observations (shared here as one of the authors of this paper): 
``The project carried out involved a 2 hour (invasive) operation to fire a 100 electrode Utah Array (later referred to as ``Braingate'') into the median nerves of the left arm for experimental purposes. With this in place 3 main experiments were performed~\citep{warwick2018neuroengineering}. 
1.~Remote neural control of a robot hand. 
2.~Extra-sensory input from ultrasonic sensory input. 
3.~Telegraphic communication between two human nervous systems.
Each experiment involved a closed-loop, cybernetic system part-human, part-machine. An important aspect of this feedback loop was the functioning of the human brain. 
Training the brain to understand stimulating pulses took time and needed to be done in short bursts as it was quite repetitive, the pulses needed to have meaning. But it was quite clear that the brain was very much able to deal with novel sensory input – in part 2 of the experiment ultrasonics represented an accurate indication of distance to objects.
Motor neural control of a remote robot hand in part 1 – person in USA, hand in England was performed by sending neural information via the internet. 
Feedback was provided by fingertip sensors on the robot hand to indicate strength of grip. 
It was quite possible to apply just enough force to grip an object. 
The person involved felt themselves to be very powerful. 
The robot hand was effectively part of their body, although it could have been any technology, not necessarily a hand.
Communications in part 3 involved sending telegraphic signals between nervous systems. 
Opening or closing a hand could be depicted. 
No false signals were witnessed and no signals were missed. 
Clearly when the same system is employed directly between two human brains it opens up the possibility of a new form of communication directly between brains, indeed between several brains at the same time if desired.''

It is fundamental to understand that the enhancement opportunities that follow from these ideas is that the technological extensions could be anthropomorphic or of whatever design.
Therefore, our approach suggests a reliable coordination between bio-plausible mathematical models (simulations) which will allow to identify the correct and functional plausible ways to hack human nervous systems and, therefore to modify our minds.

\section{Bio-plausible simulations}\label{sec:models}

The current state of the neuroscience research in neuronal microcircuits provides the background to build the model of neural circuits on the level of individual neurons and their connections, considering morphological variation and complex topology of the neuronal circuits. 
Earlier several groups of researchers published successful works with simulation of mammalian cortical columns.
\citep{traub2005single} demonstrated the model consisting of 3,560 neurons that reproduces phenomena including thalamocortical sleep spindles, persistent gamma oscillations, neocortical epileptogenesis, etc. 
\citep{breakspear2006unifying} used the mean-field theory (MFT) to reproduce the dynamics of the cortical network obtained from 19 surrogate sets of neurons taking into account the orientation of hypercolumn and structure of connections.
Later in 2012 ~\citep{zheng2012balanced} developed a model applicable to local ﬁeld potential (LFP) recordings near the soma of the layer IV pyramidal neural population and demonstrated the successful prediction of the balance between neural excitation and inhibition using extracellular recordings.
The group of Gerardo-Giorda et. al.~\citep{wei2014unification} developed the model of a Cortical Spreading Depression (CSD), which computational grid that consists of 140,208 nodes for the left hemisphere, and a mesh of 139,953 nodes for the right hemisphere and implemented the propagation of extracellular potassium waves~\citep{kroos2016geometry} as the main cause of CSD.

The research in of the spinal cord circuitry over last decades precisely described multiple projections concentrated around motoneuronal pool \citep{ampatzis_separate_2014,chopek_sub-populations_2018}. 
The modelling of these circuits explain the flexor-extensor and limbs coordination. 
The major disadvantage of the bio-visible models of spinal cord circuits is that they are unable to explain the structure and functional flexibility of the central pattern generators (CPGs). 
For this purpose several research groups implemented modelling approach with some fruitful results.

The identification and modeling of the functional microcircuits within the spinal cord segments where initially were identified the basic circuits including reflex arc. 
Later several groups~\citep{rybak2006modelling, shevtsova_organization_2016}
described rhythm generating and pattern formation circuits. 
Another modelling approach 
to the spinal circuitry was taken to create a 3D model of the spinal cord segment accounting the role of the electric stimulation in activation of the different afferents, along with interneuronal and motoneuronal pools.

\subsection{Neurocomputational models}\label{sec:software}

\textbf{Definition 1 }\textit{Assuming that the  unit of the information transfer in nervous system is the action potential or the spike and it's average absolute refractory period is $1\si{\milli\second}$; thus the computing of action potential generation 
processes must be done during the action potential absolute refractory period of $1\si{\milli\second}$.}

In order to provide optimal integration neurocomputational models with the neuronal systems or novel robotic systems the criterion of real-time or semi real-time their operation have to be identified (Definition 1). In our opinion, the key requirements for the simulation of neuronal circuits includes: (1) the real-time processing of the whole neuronal activity of the part of simulated biological system and DAC-ADC (digital-to-analog and analog-to-digital) conversion, (2) relative bio-plausibility limited by the requirement 1, (3) neurocomputational models should be wearable or even implantable.

The balance between an acceptable level of bio-plausibility or details in the model and time processing is usually not trivial question. There are several efficient models of neurons to operate real-time still being close to neurobiological processes. We consider them in more details in the next subsection.

\subsubsection{Spiking neurons and NNs}\label{sec:snns}

As biological experiments show, for neurons at various sites of the nervous system, a typical property is oscillatory activity (for example, hippocampal neurons, neurons of the cortex and some interneurons). 
Being expressed in fast and slow time scales, such activity can be chaotic.
One of the first significant results in neurodynamics, which are important to our days, is the Hodgkin-Huxley model, which describes the principles of generating an action potential in a neuron. The model demonstrates two dynamic modes: an excitable mode and a mode of generating a periodic sequence of pulses \citep{chen2020structure}. Based on the Hodgkin-Huxley model, a large number of mathematical models of neurons have been developed \citep{calim2018chimera}, which are used as basic elements for the construction of models of large neural networks and their implementations~\citep{mishchenko2017instrumental}. 
One of the first reduced small-size Hodgkin-Huxley models can be called the Fitzhugh-Nagumo radio engineering model (1961), which is convenient not only for analytical research, but also for radio engineering prototyping. 
This model can have coexisting attractors, i.e. the considered system is bistable.
Depending on the initial conditions, the trajectories can converge to one of two equilibrium states - stable or unstable. 
The basins of attraction are separated by separatrices~\citep{nagumo1962active}.

The model of Izhikevich \citep{izhikevich2004model} is also simplification of the Hodgkin-Huxley model \citep{hodgkin1952}. 
The Izhikevich model does not simulate the ion flow across the membrane instead calculating the probability of ion channels activation/inactivation. 
In the ESRN (even simpler real-time neuron) model \citep{leukhin2020even}, the membrane potential is a simplified equation of the sum of synaptic weights, current leakage, and a noise generator. 
This simplification provides the real-time calculation of thousands of neurons.
In the paper of Izhikevich \citep{izhikevich2004model}, the comparison of neuron spiking model was presented, with the most efficient model ``integrate and fire'' \citep{smith2000fourier}, but this model does not ensure good bio-plausibility whereas the Izhikevich model shows better bio-plausibility although requires higher computational power. 
The ESRN model is more efficient than the LIF (leaky integrate and fire) model but shows a similar low bio-plausibility level. 
In our comparison \citep{leukhin2020even}, ESRN and LIF fit the real-time requirement.

Subsequently, more detailed models were proposed that consider the dynamics of other ion currents and possess qualitatively new dynamic modes. 
The addition of a calcium current to the model leads to the appearance of a resonant bifurcation with a solution branching at the critical point with another periodic solution of the same period. This method allows obtaining a burst mode in the model, in which, in response to a suprathreshold excitation, the oscillator generates not one, but several pulses (groups of spikes) \citep{izhikevich2004model}. Later, other models with the presence of burst modes have been proposed (for example, \citep{gonzalez2015biologically}). 
For the first time, the model of neural bursting activity was proposed in the work of Hindmarsh and Rose \citep{hindmarsh1984model}. 
The Hindmarsh-Rose model demonstrates bistable behavior, i.e. the coexistence of a stable limit cycle and a stable equilibrium state. 
Also, the Hindmarsh-Rose model is capable of not only regular, but also chaotic generation of spikes \citep{dmitrichev2018nonlinear}. 
It is also worth noting the radio-engineering implementation of the model \citep{romo2019synchronization}.

Neural coupling occurs due to the so-called synaptic connections, which have a complex spatial architecture and provide a diverse nature of inter-neuronal interactions. One of the effects of synaptic communication is the forced synchronization of the receiving neuron with the transmitting one \citep{gerasimova2017simulation}. Neuron models are linked together using unidirectional electrical connections to mimic inter-neuronal interactions. 
In 2011 an active optical communication channel between neural generators using fiber-optic lasers to simulate synapses \citep{pisarchik2013experimental} was proposed and implemented. 
This system allows transmitting signals between generators and using the laser as a synapse to control the dynamics of the active communication channel. 
With a relatively simple dynamics of a neuron-like generator, such a connection provides a wide variety of synchronous, quasi-synchronous, and chaotic regimes. Such a variety of dynamic modes of the system indicates plasticity - flexibility of the proposed optical synapse. 

The plasticity effect is inherent in almost all neuronal cells. Synaptic plasticity was first considered as a mechanism of higher cognitive functions based on theoretical analysis in 1949 \citep{hebb1949organisation}. At present, many models and variations of spike timing dependent plasticity mechanisms are known, mathematically designed, analyzed \citep{houben2020calcium}, and the electrical implementation of such a process has also been demonstrated \citep{cameron2005spike}. The variety of models of synaptic plasticity reflects the importance and relevance of this phenomenon in describing information processing and brain functioning. 

When studying collective processes in ensembles of neuron-like elements, the main issue is the ability of such systems to form spatiotemporal structures - patterns.
This problem has been widely studied in connection with various aspects of both neurodynamics and other areas of nonlinear physics. 
The processes of the formation of stationary spatial structures in bistable systems~\citep{tyukin2019simple}, synchronization of the ensembles of self-oscillating systems~\citep{iudin2016simple}, synchronization of chaotic systems~\citep{vaidyanathan2015adaptive}, the phenomenon of cluster formation~\citep{baldassi2015subdominant}, the formation of chimeric states of dynamical systems (for which synchronous and asynchronous modes coexist in one homogeneous system in different areas of space) have been studied \citep{kasatkin2019itinerant}. 
The results obtained allow estimating the regions of the parameters of the existence and stability of certain dynamic regimes, the conditions for their formation and the main dynamic, statistical and informational characteristics. The tasks of saving and transforming information by multielement neuron-like systems lead to the study and modeling of so-called patterns of neuronal activity (packets of nerve impulses) with given spatial and temporal characteristics.

The associative memory model, first proposed by Hopfield in the 1980s, opened a new period in the application of ideas and methods of statistical physics to neuroscience \citep{hopfield1982neural}. 
The theory presented by Hopfield was confirmed experimentally in studies on monkeys. 
When the monkey was recalling the first stimulus, neurons were found to exhibit increased selective neural activity during the entire delay, which is usually a few seconds. 
Hopfield's model also explains the formation of short-term couplings between states in a learning sequence. As shown experimentally, neurons display a group of elements that have adjacent positions in the training set. In this case, the elements can be completely uncorrelated with each other. This can be explained by Hebb's modification of neurons, which originally represented neighboring elements. Various models of multielement neural networks \citep{hopfield1982neural} based on Hopfield networks allowed simulating Hebbian learning rules \citep{hebb1949} to construct theoretical models of the phenomenon of associative memory. On the basis of theoretical studies of this phenomenon, neural networks of the perceptronic type were built, which laid the foundation for the development of neurocomputer \citep{grewe2017neural}.

Later, neural networks have been developed in the form of arrays of dynamical systems (cells) - cellular neural networks or cellular nonlinear networks (CNN) \citep{huang2020asymptotically}. These neural networks have been designed as integrated circuits. Among the numerous applications of cellular neural networks, the most common is image processing. These systems are also of interest as an object of research in nonlinear dynamics, since there is a transformation of dynamic structures and the emergence of a wide range of dynamic states \citep{aouiti2016neutral}.

In the area of nonlinear dynamics, complex effects associated with the formation of new structures are widely studied with a special attention paid to the problems of assigning a certain configuration in large neural networks and identifying the oscillatory properties of neural networks.
As an example, we can refer to the phase model of the Kuramoto neural network, where the models of phase oscillators \citep{greenwood2016kuramoto} interacting in a certain configuration were used as neurons.
This model is characterized by the appearance of phase clusters - the existence of in-phase and anti-phase oscillations in space. 
Kuramoto phase generators are used for image segmentation, where each selected feature is encoded by an ensemble of synchronous generators, as in biologically plausible systems.
Such models are designed to perform color  and object segmentation using the phenomena of synchronization between generators in image processing due to the operation of a multicore neural network \citep{novikov2015oscillatory}. 
The problem of synchronization of Kuramoto oscillators with non-identical generators (non-identical natural frequencies) was studied in \citep{sun2020fixed}. 
Not only cluster, but also chimeric solutions were demonstrated on the model of identical Kuramoto-Sakaguchi phase generators \citep{ichiki2020diversity}. 
It was shown that a certain configuration of the cluster and the natural frequency of the generators, which provide the ability to restore the cluster synchronization configuration after the disturbance of the states of the oscillators, i.e. asymptotic stability \citep{menara2019stability}. 
The applied aspect of the construction of artificial neural networks is associated with the prospects for the construction of new information, adaptive and self-organizing systems.

In our opinion, a qualitative breakthrough in the development of spiking NNs can be based on the concept of ``blessing of dimensionality'' \citep{gorban2019unreasonable}. According to this concept, certain methods can work much better in high-dimensional spaces than in low-dimensional ones. It is well known in the brain sciences that small groups of neurons, or even single neurons, play an important role in cognitive functions. Recently, it has been shown \citep{gorban2019unreasonable} that ensembles of noninteracting, dynamically simple, but at the same time high-dimensional neurons turn out to be an effective tool for solving essentially high-dimensional problems that often can’t be solved by standard analysis. The new fundamental approaches dealing with high-dimensional data should be based on the hardware implementation of artificial neural networks provided by the development of electronics and its new element base. In recent report \citep{makarov2021towards}, the mathematical problem of processing complex high-dimensional data in AI systems has been for the first time projected onto new approaches to the implementation of learning and associative memory in relatively simple neural network architectures based on memristive electronic devices with rich internal dynamics. These approaches are correlated with the well-known ``revolution of simplicity in neurosciences'', which will allow to reproduce in hardware the complex cognitive phenomena and create prerequisites for the construction of the first prototype of an artificial hippocampus in the medium term.

\subsection{Hardware implementation}\label{sec:hardware}

Significant progress in the capabilities of computing systems and information technologies that we have observed over the past decades was provided by the step-by-step reduction in the size of transistors fabricated using CMOS (complementary metal-oxide-semiconductor) technology in accordance with the well-known Moore's law \citep{moore1965cramming}. The number of transistors on a chip has doubled approximately every 2 years with an exponential increase in the speed of microprocessors. 
However, the slowdown of Moore's law, caused by a number of reasons, does not allow microelectronics to continue to follow the previous trend. 
First, the performance and speed have been limited for over 10 years by heat dissipation in highly integrated circuits \citep{sutter2005free}. 
Second, the characteristic size of transistor has already come very close to the fundamental physical limit of the order of 2-3 nm, at which quantum phenomena and uncertainties make the operation of transistor unreliable in traditional circuits \citep{khan2018science}. And, worst of all, the difference in performance between information processing and storage units has increased dramatically, so data movement between these units becomes the main reason for high power consumption and latency in the traditional von Neumann architecture \citep{horowitz20141}. Widely known as the bottleneck of the von Neumann architecture, this problem will be exacerbated in data-intensive applications such as machine learning. To mitigate these challenges, new materials, devices and computing architectures are currently being actively investigated. They are intended to complement and, possibly, replace conventional devices and circuits based on CMOS technology.

A striking example of such materials and devices are nonvolatile memories based on the phenomenon of resistive switching (RS) \citep{ielmini2015resistive}, which in 2008 \citep{strukov2008missing} were associated with the memristor as the fourth passive circuit element, reversibly changing its resistance under the flow of electric charge \citep{chua1971memristor}. Unlike traditional memory devices, which use electric charge to store information, RS devices store information in the form of a resistance value, the change of which is determined by the rearrangement of the atomic structure in thin dielectric or semiconductor films of nanometer thickness under the action of electric field / current. The anionic-type RS phenomenon (Valence-Change Memory – VCM) is manifested in metal-oxide-metal (MOM) structures, which are most compatible with the traditional CMOS technology, and is related to the reversible formation / destruction of conducting channels (filaments) in the oxide film due to a combination of oxygen ion migration, reduction and oxidation processes. Such MOM-structures are the main constituents of ReRAM or RRAM (Resistive Random-Access Memory) memory devices.

ReRAM is attracting great interest as a universal non-volatile memory that combines the characteristics of existing types of memory, including SRAM (Static Random-Access Memory), DRAM (Dynamic Random-Access Memory) and read-only storage devices in the form of SSD (Solid State Drive) or HDD (Hard Disk Drive), thanks to high switching speed, low power consumption, high reliability and scalability \citep{ielmini2015resistive}. Particularly attractive is the possibility of achieving high density and three-dimensional integration of ReRAM arrays due to the simple two-terminal structure of the memristor and the locality of RS phenomenon (down to nanoscale) \citep{kim2018scaling}.

In addition to applications directly as non-volatile memory, significant efforts in recent years have focused on using ReRAM devices and arrays to perform computations at the storage location, known as in-memory computing \citep{zidan2018future}. This approach fundamentally solves the bottleneck problem of the von Neumann architecture, eliminating the need to constantly move data between the processor and memory units. Moreover, an array of ReRAM devices in a cross-bar topology is ideal for hardware implementation of neural networks \citep{xia2019memristive}, naturally realizing the vector-matrix multiplication (VMM) operations based simply on Ohm's and Kirchhoff's laws. Since VMMs are the most used operations in typical neural network algorithms, their hardware implementation in the cross-bar topology makes it possible to increase the performance and speed of neuromorphic computing systems by orders of magnitude. 
It is important to note that rich dynamics of memristive devices can be used to accurately simulate many biological processes, including synaptic and neuronal functionalities described in Section 3.1.1 and critical for learning and memory. This will enable more efficient neuromorphic systems capable of operating at the interface with living neural systems \citep{mikhaylov_neurohybrid_2020}.

All described applications of memristive devices as an element base for prototypes of new generation information and computing systems have become the subject of numerous publications in recent years (more than 2000 publications in 2020 on the Web of Science database). In the same 2020, more than 10 high-quality reviews were published, in which the achieved characteristics of neuromorphic computing systems based on arrays of memristive devices of different sizes were analyzed in detail and a roadmap for the development of this research field was presented \citep{zhang2020neuro}. The prototypes of memristive neural networks already demonstrated so far are comparable in performance with existing neuroprocessors based on traditional digital electronics and specialized architectures (ASICs), such as TrueNorth (IBM), Loihi (Intel) and Tianjic (Tsinghua University), and are 2-3 orders of magnitude ahead of the latter in energy efficiency. Within the next 5-10 years, the creation of general-purpose memristive neuroprocessors is expected. At the moment, the main efforts of researchers and engineers are focused on the co-optimization of memristive materials and devices in accordance with the requirements for specific emerging systems and technologies. Three-dimensional integration of memristive devices is a promising way to the development of future ultra-large neuromorphic integrated circuits that can approach the capabilities of the human brain \citep{veluri2021high}. An important role on this way is played by the implementation of a systematic approach to the development of the entire chain of computer-aided design (CAD) tools from devices to algorithms and hybrid software-hardware simulation systems discussed in Section 3.3. It is the high energy efficiency and unique scalability of memristive systems that make it possible to take the next step from memristive neuromorphic computing systems to neurohybrid systems based on the symbiosis of artificial electronic systems and living neural systems for solving relevant problems of artificial intelligence, robotics and medicine \citep{mikhaylov_neurohybrid_2020}.

At the moment, the best-known prototypes of close-loop neurointerfaces, which will be discussed in detail below \citep{boi2016bidirectional} use complex mathematical models and software on the side of an artificial system, implemented on high-performance computers or specialized neuroprocessors using traditional electronics. The most recent hardware example of such a neuromorphic system \citep{sharifshazileh2021electronic} is customized for the specific task of recognizing high-frequency oscillations in pre-recorded intracranial EEG (iEEG) signals from patients with epilepsy and is a spiking neural network using 4 cores of 256 LIF neurons each and synapse models with simple first-order dynamics. The preprocessing of iEEG signals and their transformation into spike sequences are also implemented on a single chip with neuromorphic cores, but all system elements are controlled by FPGAs on a standard test board. Obviously, the traditional element base cannot ensure the achievement of high requirements for compactness (miniaturization) and energy efficiency of artificial neuromorphic systems for their subsequent direct and safe interfacing with living neural networks.

The first examples known from the literature, in which memristive devices and arrays are used to process bioelectric activity, just confirm the fact of communication of electronic and biological systems through discrete memristive devices \citep{serb2020memristive} or do it in isolation from the living systems themselves (for example, in recent reports \citep{liu2020neural}, memristive chips process signals of neuronal activity taken from public databases).

The greatest progress in memristive neurohybrid systems has been achieved in recent work \citep{shchanikov2021designing} that demonstrates the world's first bi-directional adaptive neurointerface using advanced solutions in the field of memristive electronics and neuroengineering. To create an electronic subsystem, an optimized technology of metal-oxide memristive microdevices is used \citep{mikhaylov2020multilayer} together with a classical neural network architecture such as a multilayer perceptron. From the side of living system, the culture of hippocampal neurons grown on a multi-electrode array is used with functional connections between groups of neurons spatially ordered by a microfluidic chip. The memristive network is for the first time used not only to solve the problem of nonlinear classification of the spatio-temporal response of a cellular culture to electrical stimuli, but also to control its functional state. Namely, the output signals of the memristive network correspond to different stimuli and are used for adaptation of stimulation to restore disturbed functional connections in neural culture. All developed solutions are implemented in the form of electrical circuit prototypes, software packages and are suitable for subsequent monolithic integration within the framework of a new concept of a neurohybrid chip and a roadmap until 2030 \citep{mikhaylov_neurohybrid_2020}.

\subsubsection{Memristive synapses and neurons}

Memristive devices have sparked particular interest in the hardware implementation community because of their promise to create extremely energy-~\citep{Strachan2011} and area-efficient~\citep{Khiat2016} neural components. 
The vast majority of work in this area has thus far focused on building memristive synapses because of the overwhelming numbers of synapses in neuron-based computing systems; typically outnumbering neurons by 100-10k to 1. 
Memristive synapses take advantage of the very direct correspondence between a basic memristor (a modifiable, non-volatile resistance) and a basic synapse (a modifiable, non-volatile weight): resistance is mapped onto the concept of a weight. This allows extremely simple, two-terminal, passive components that can be aggregated into ultra-dense crossbar arrays \citep{Pi2019} and stacked atop each other in 3D \citep{Sun2019} to embody potentially very large numbers of artificial synapses \citep{Sebastian2017}.

Naturally, biological synapses are far more complicated than the idea of simple weights obeying simple learning rules might suggest. There has been a large body of work attempting to create ``high horsepower'' memristive synapses, where the physics of memristive devices are exploited for emulating more complex behaviours observed in full-blown biological synapses. 
These include short-term effects \citep{Ohno2011}, a natural propensity for STDP \citep{prezioso_self-adaptive_2016}, soft boundaries \citep{Serb2016a}, metaplasticity \citep{Zhu2017}, stochasticity \citep{Vincent2015} and more. It is still unclear either to what extent this shift of ``computational burden'' down to the device physics level can progress, or what the optimum set of behaviours to ``physicise'' is.

Finally, we note that the ability of memristors to exhibit non-linearity (e.g. thresholding), opportunities to use them as implementations of neuronal membrane segments have also been investigated. A classical example is the ``neuristor'' \citep{Pickett2013}, showing how a ``transmission line'' of memristor pairs can be used to regeneratively propagate a neural spike-like waveform down its length. In a similar vein memristive weighing can be used to implement dendritic structures \citep{Zhanbossinov2016}. The confluence of active membrane circuitry, artificial synapses and multiple families of memristive devices, each exhibiting their own electrical properties raises the alluring prospect of memristor-only neural networks. This is of particular importance because unlike transistors, memristors do not consume Silicon real estate and can be stacked atop each other in principle indefinitely many times (requiring only regular layers of ground and power planes to keep the active membranes functioning).

It will be interesting to see what the ultimate horizon of this technology turns out to be and how much better it can perform over standard, exclusively CMOS-based circuitry. 
Moreover, it will be interesting to observe how much of the neuron's functionality will be transferred to the memristive device. 
For example, combining the axon-like regeneratively propagating wave properties of the neuristor with dendritic and synaptic weighting implemented directly as memristive devices it is in theory possible to replace the entire neuron with such components, eliminating the need for transistors altogether for everything but diagnostic purposes. 
However, balancing and controlling the activity within such networks appears extremely challenging without the explicit control of transistor-based circuits (how would one perform the initial programming of the network's weights?). 
However, assuming the problem of controllability is solved (perhaps through some futuristic ``self-evolving from 0'' neural network) the promise of fully back-end integrable neurons seems extremely alluring: it would in principle allow 3D stacking in the 10s of neurons atop each other based on a fabric of power supply planes sandwiching memristor-powered circuits. Such systems, having no access ports for explicit reprogramming might also prove very resilient to hacking, providing another enticing potential benefit.

\subsubsection{Organic memristive systems}

As it was already described, memristive devices are considered as very perspective elements for the effective brain-computer coupling, as they combine memory and processing functions, mimicking, therefore, essential synapse properties. 
The most of memristive devices now are based on inorganic (mainly, metal oxides) materials. 
This choice of materials is mainly due to the fact that it is much easier to integrate inorganic memristive devices into existing CMOS technology, that is the basis of the electronics now \citep{mikhaylov_neurohybrid_2020}.

Nevertheless, organic materials are also widely used for the memristive devices realization, because these devices can have several important advantages, such as low weight, flexibility, low energy consumption, bio-compatibility, low operation voltages, capability to self-organization in complex 3D systems, etc.
Briefly the overview and comparison of properties of organic and inorganic memristive devices are presented in \citep{erokhin2020memristive} and in more detail in \citep{erokhinfundamentals}.
Several organic materials were used as active layers of memristive devices. We will not consider here all these materials and address readers to two mentioned papers \citep{erokhin2020memristive,erokhinfundamentals}. Here we discuss only two examples: memristive devices based on polyaniline (PANI) and on parylene.
PANI-based memristor is the most studied organic memristive device \citep{Erokhin_2011}. It was designed and realized exactly for mimicking properties of synapses: memory effect (hysteresis) and rectification (unidirectional signal propagation). 
It was shown that the device can be used as a key element in artificial neuron networks \citep{emelyanov_first_2016}, logic with memory \citep{erokhin2012organic}, nervous system mimicking circuits \citep{erokhin2011material}, oscillators \citep{Smerieri_2008}, etc. 
It has been also shown that these elements can couple directly live nervous cells from the rat cortex, demonstrating properties absolutely similar to those of the natural synapses \citep{juzekaeva_coupling_2018}. 
The system has effectively demonstrated a frequency driven long- and short-term potentiation and depression \citep{battistoni2019frequency}. 
The use of these devices as synapse mimicking electronic elements has demonstrated the possibility of unsupervised learning according to the STDP algorithm, what has allowed to realize electronic circuits, demonstrating classic conditioning \citep{prudnikov2020associative}. 
However, the main advantage of such systems is the capability of organic molecules to self-organization into 3D systems \citep{erokhin2012stochastic}. It is difficult to imagine that inorganic systems, fabricated with current electronic technologies will reached an integration level that the brain has (1015 synapses \citep{erokhin2010organic}). Instead, the use of specially synthetized block-copolymers resulted in the realization of 3D systems with the level of memristive devices integration was about 1011 $cm^3$. 
Parylene based memristive devices are also very promising elements for neuromorphic applications. This material is widely used in, in particular, sensor applications due to easy fabrication process, light weight, flexibility and bio-compatibility  \citep{khodagholy2011highly}. Several memristive systems were fabricated using this material  \citep{chen2019low}. The mechanism of the resistance switching was attributed to the metal filament growth in parylene \citep{minnekhanov2019resistive}. Even if resistance switching potentials in this case are higher than in the case of PANI-based memristive devices, these systems have one important advantage: the fabrication technology can be easier integrated into existing technological processes. It is to note that these devices were effectively used for the realization of circuits, allowing classic conditioning according to the STDP algorithm \citep{minnekhanov2019parylene}.
Summarizing, organic memristive devices must be seriously considered as key elements of cyber punk revolution. They can play a role of synapses in external electronic circuits and can provide direct synapse-like connections between nervous cells in living beings.

\subsection{Hardware-software simulation}

The transition to a new level of Cyborg 3.0 requires the use of modern and effective development and design technologies. The devices considered in this paper are complex real-time closed-loop embedded systems that have a large number of internal interconnected components and interact with the objects from the external environment through the sensory and motor parts. 

Modeling is an essential part of the development process of such systems and at the different stages of this process, various types of software and hardware models and their combinations aimed at solving different tasks are used. 
Currently, the most advanced achievements in this field are accumulated in the methodology of systems engineering. According to that methodology a set of models of different levels of the structural and functional hierarchy is created during the development of the systems and it is called a ``digital twin''. A ``digital twin'' made it possible to test and optimize a system under development (its architecture, structure or other parameters) and predict its operation throughout the life cycle.

For a wide range of tasks in this area, general-purpose numerical modeling applications and languages (such as MATLAB, Scilab, GNU Octave, Mathematica, etc.) can be used, as well as electronic circuit simulators based on SPICE (such as LTspice or PSpice, etc.). However, these software tools are not able to cover the most complex tasks related to modeling the activity of biological neural networks and neurohybrid systems. Moreover, software modeling of such systems requires significant computing resources, so it is necessary to use software tools that are able to accelerate the simulation process through the use of hardware accelerators. 

Currently, there are several technologies that may be used for such hardware-software simulation. On the software side, there are programs such as GENESIS, NEURON, MOOSE, etc. These programs provide functionality for creating models of different types of neurons (e.g. integrate-and-fire neurons with current or conductance based synapses, Izhikevich and Hodgkin-Huxley models, etc), synapse models, including short-term plasticity and different variants of STDP, spiking or traditional non-spiking models of neural networks, multiscale modelling of local networks and brain areas and much more. In the article~\citep{Birgiolas2018} one can find an overview of some of the most widely used software applications for simulating models at various levels of biological detail with links and detailed descriptions. This software supports various hardware acceleration options from more affordable multi-core CPUs and GPUs (NVidia CUDA/AMD Stream, OpenCL, OpenMP, etc.) or FPGA to Intel’s Loihi Architecture, SpiNNaker, BrainScaleS, MPI-parallelism in computer clusters and super computers (like IBM BlueGene). Such hardware-software simulation systems can be orders of magnitude faster than the simple software execution on an ordinary personal computer. A comparative overview of such technologies is available in~\citep{tikidji2017software}.

The use of specialized hardware-software simulation environment can be a part of the process of studying complex systems, allowing you to accelerate numerical experiments, and also it can be directly a part of the process of system engineering. One of the most important and difficult stages during the development of such systems is hardware-in-the-loop (HIL) simulation~\citep{Shchanikov2021HIL}. HIL-simulation methodology is used in the cases of creating systems that are parts of other more complex systems (in this case, living biological systems). 

This makes it possible to improve the quality of development, making a system more flexible and adaptable to changing parameters of the external environment. 

HIL-simulation also allows testing, validation and performance evaluation, answering the question whether the system is able to process in real time all the information from the objects with which it interacts \citep{yang2018control}. 
HIL-simulation is safer and cheaper than testing systems in-vivo and makes it possible to provide scenarios for safe handling of emergency situations (for example, for extraneous signals). 
This is a very important and relevant aspect, since for this field of research, the behavior of objects of the external environment has dynamical nature and its modeling is a complex scientific task itself.

\section{Biological interface}\label{sec:mbsi}

We can see several examples of the creation of interfaces connecting machines and biological systems, from classic engineering biohybrid attempts \citep{rochford2020bio} to synthetic biology approaches \citep{yarkoni2015solving}, such technologies have established successful bonds between the natural and artificial realms. Despite of the currently dominant neural and brain approaches \citep{donoghue2002connecting}, there are several other ways to connect machines and biological systems. Such bottom-up perspective allowed to pave the way for a more complex generation of interfaces \citep{fletcher2016bottom}. At the crossroad, the experimentation of biobots or living machines \citep{kamm2014creating} opened new connections between machines and living systems, even to the birth of xenobots \citep{ball2020living}. 
The other successful way to connect machines and biological systems has been the field of unconventional computing \citep{adamatzky2007unconventional}, or even wetware computing \citep{dennett2014software}. In relation to wetware computing, must be related here the impressive results of the EU FP7 Project ``Physarum Chip'', in which Andrew Adamatzky (University of the West of England, Bristol, UK) and Theresa Schubert (Bauhaus-University Weimar, Germany) constructed logical circuits that exploited networks of interconnected slime mold tubes to process information \citep{adamatzky2014slime}. On the other hand, the hybrid biochip of Prof.Warwick allowed that rat neurons operated a computer chip and also a full robot~\citep{warwick2011experiments}. 
In that successful experiment rat neuron cells were grown on an \emph{in vitro biochip} which was running via Bluetooth a robot (due to technical survival requirements of the \emph{in vitro biochip}). 
Following Prof. Warwick revision notes: ``For the rat brain robot experiment it was possible to witness (under the microscope) new neural passageways being formed in a relatively short space of time. This amounted to connections between neurons formed by axons and dendrites thickening/strengthening over time, effectively a physical and mental response at the same time. Such pathways then became more likely to be employed next time stimulating pulses were received. The neurons involved were acting as communicating engines (nothing more) either between other neurons, or from sensory inputs or to motor outputs. Learning merely dictated which pathways would be more likely to be employed subsequently. Neurons took on specific roles because of their physical location in the reconstructed brain.''
The main challenges of such systems are the functional compatibility, or the conceptual blending bending the limits of the natural and artificial entities, something with also legal related issues \citep{hockfield2019age}.
From a taxonomic and system perspective, we could define four main  machine to biological systems interfaces: cell-machine, plant-machine, animal-machine, and human brain-machine interfaces. 
Cell-machine connections have intensively studied at gene regulatory level~\citep{zillig2021cells}, a very specific domain application, while plant-machine studies are more related to conceptual explorations towards the still understudied field of plant cognition~\citep{calvo2007quest}.
Nevertheless, some cutting-edge research times, like MIT Media Lab's Fluid Interfaces Group are exploring such possibilities, announcing ``cyborg botany'' \citep{sareen2019cyborg}. Animal-machine studies have been mainly devoted to pave the way for human-machine neural implants \citep{nuyujukian2011monkey}, although there are studies that are truly interested on animal-machine specific interactions \citep{savage2000animal}. Some of these interests are from a military perspective, like stentrodes implanted into sheep by a DARPA funded project \citep{graham2016minimally}, something affected by ethical concerns.

\section{Brain-computer interface, BCI}\label{sec:bci}

%%%
The brain-computer interfaces (BCIs) or the neural interfaces are the hard\-ware-software systems for the functional interconnection between a biological object and a machine, i.e., for direct connection of a computer or some digital intelligent control system with the nervous system, first and foremost with the brain \citep{hramov2021physical}. Implementing BCIs implies a new channel for communication of the person with external devices and/or control them, without the interference of peripheral nerves and muscles, instead of using keyboards, mice, joysticks, and other specific equipment, such as eye trackers following a view direction, etc. BCIs now provide the closest technological integration between a living object and a machine, creating a smooth information channel between our brain and a digital platform by interpreting our mental intentions or assessing changes in psychophysiological states.

\subsection{Brain signals for BCI developing}

The central core of any BCI is an intelligent system that enables to classify brain states in real-time according to recorded brain activity due to either spontaneous physiological processes or external stimulation. 
The BCI then transforms the revealed features of the brain states into control commands for external applications (exoskeleton, bioprosthesis or wheelchair control, attention or emotional state monitoring, etc.). 
In the modern BCIs, the brain activity is recorded usually by invasive methods such as cerebral cortical registration (CCR) and electrocorticography (ECoG) or non-invasive techniques as  electro- (EEG) and magnetoencephalograpy (MEG), functional near-infrared spectroscopy (fNIRS), etc., which to some extent reflect the functions of the central nervous system. 
Invasive BCI provides a much better quality of registered neural signals. 
The invasive BCI is based on CCR and ECoG recordings from single brain cells, or multiple neurons \citep{lebedev2006brain}. 
Due to surgical risks and ethical issues, invasive BCI technology is only applied in humans in the case of medical indications. 
At the same time, there are many examples of invasive BCI realizations with animals and first of all with monkeys and rodents. 
In contrast, non-invasive BCI techniques have the great advantage of not exposing the subject to the risks of brain surgery but provide communication channels of limited capacity \citep{hramov2021physical}.

\subsection{BCI classification}

%Various types of commands \todo{this is not clear controlling instructions? or just control} given by BCI operator 
Patterns of brain activity triggered by the BCI operator purposefully or spontaneously can be used to organize information stream from the brain to the machine through the neuronal interface \citep{zander2011towards}. 
Active BCIs use changes in the brain activity, directly and consciously controlled by the BCI operator, regardless of external events, for control commands. 
Reactive BCIs detect and classify the brain response (for example, evoked potential) to external stimulation (visual, auditory, tactile, etc.) for control commands. Passive BCIs analyze the current brain activity of the user without any target monitoring to obtain information about the actual brain state, for example, attention, switching activity, emotional state, etc.

\subsection{BCI functional scheme}

\begin{figure}[ht]
\centering
\includegraphics[width = 1.\textwidth]{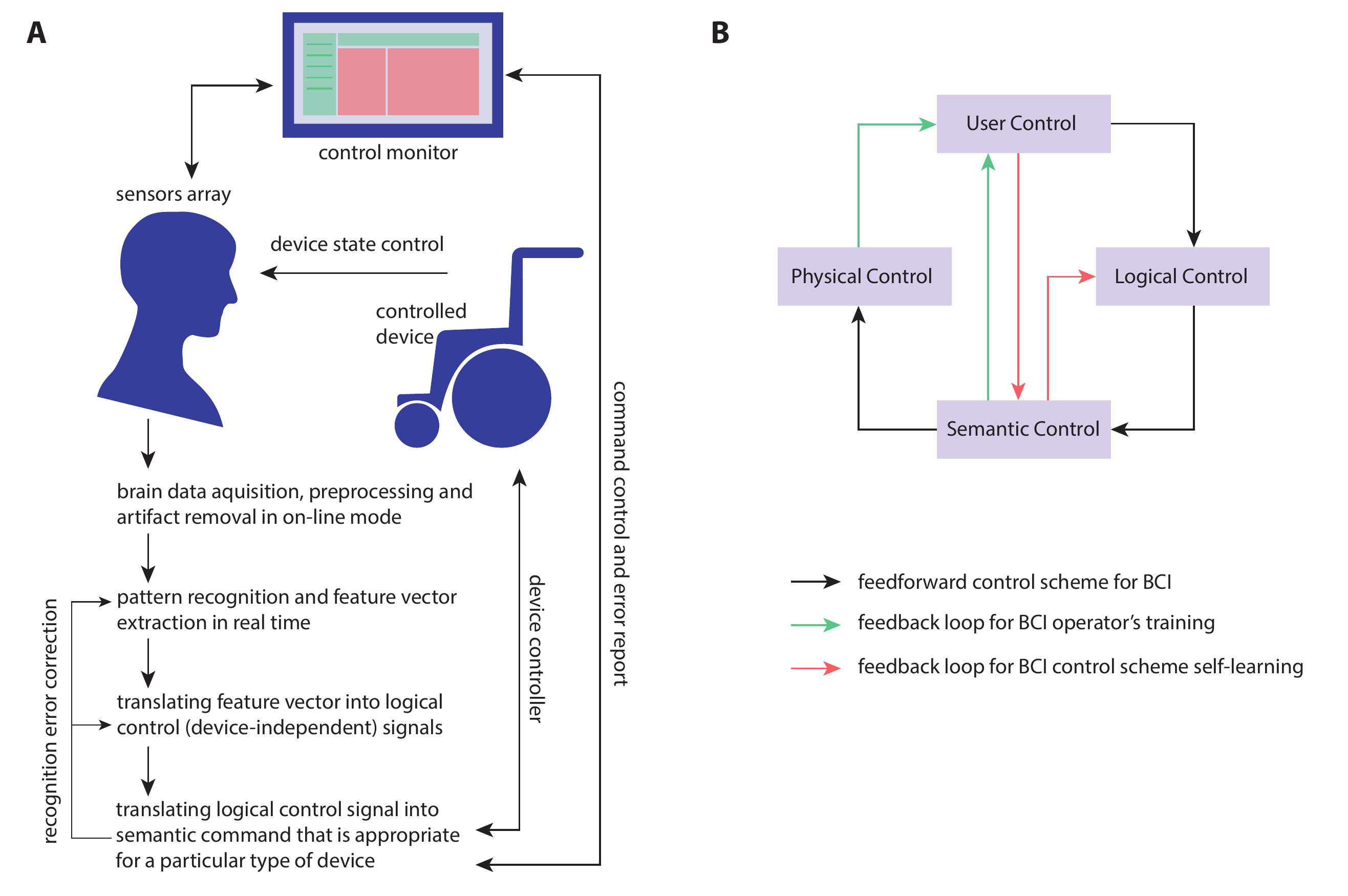}
\caption{(a) General BCI model with main functional blocks: registration of multichannel data on brain activity; intelligent processing of the obtained data and identification of characteristic patterns in real time; translating feature vector into device-independent logical command; transfer of control commands to the interface hardware. Implementation of biological feedback to monitor the command execution of, training the operator to call necessary mental condition or exposure to it depending on his diagnosed condition. (b) Closed-loop interaction of functional blocks of BCI control system and BCI user leading to simultaneous self-training BCI operator for her/his mental control and learning and adapting BCI control algorithms} 
\label{fig:BCI1}
\end{figure}

Figure~\ref{fig:BCI1}a schematically shows the functional scheme of an active BCI, in which a person (operator) controls a machine (for example, a wheelchair) through a series of functional components of the control system. 
The real-time implementation of information flow between brain and BCI is illustrated in Figure~\ref{fig:layered_technologies_stack}. 
The processing brain signals and the generation of control commands  include several successive phases, namely, data collection, their pre-processing (including the removal of recording artifacts) to prepare signals in a suitable form for further processing, forming a feature vector for identifying discriminant information in recorded signals, classification of signals based on the selected feature vector, and finally a control phase for transforming selected patterns of the brain activity into meaningful commands for any external device, such as, e.g., a wheelchair, or setting the direction of the cursor movement on a monitor screen. 

It should be noted that the pattern recognition of the brain activity of interest is significant for the BCI control commands formation. In terms of the machine learning theory, the BCI should form a feature vector determined by the peculiarity of the registered pattern of brain activity. For example, if the operator controls brain activity in alpha (8--13 Hz) and beta (14--30 Hz) frequency bands, the BCI will form a feature vector containing the EEG powers in these specified ranges. The feature converter translates the generated feature vector into a logical control signal independent of any semantic knowledge about the device being controlled and its control methods. The control interface converts a logical control signal into a semantic one already defined by a specific controlled device. For a modern active BCI driven by mental intents, this stage is universal for any brain-machine interaction.

The semantic set of commands can be not only static but also dynamic and even synchronized with the state of the controlled device. In this case, the dynamic set of commands can be formed directly in a menu, from which the BCI operator can select a required action from a large number of commands. The advantage of this dynamic approach is that the operator can set a lot of semantic control commands with a limited set of logical commands. For example, two logical commands are enough to navigate in an arbitrarily large and complicated hierarchical command menu. The result of forming a semantic command can be displayed on the screen to visualize the interpretation of the logical command in semantic. For example, the display may show a menu in which the mental navigation of the BCI operator occurs. Using the control display, it is also possible to implement biological feedback during the operator's training. 

The external device or software is managed when a physical command is formed on the base of semantic commands. Generally, if an intelligent system of pattern recognition and formation of a feature vector and a logical signal is adaptive, then a registration error forms a feedback loop that can modify the adaptive control system. It is important to note that effective BCI operation is impossible without feedback between the BCI intelligent system and the BCI operator.  For this, BCI is often completed with a control display that shows the results of interpreting the operator's mental commands in an understandable (semantic) format by the intelligent control system, both for monitoring the BCI operation and providing biological feedback. This is necessary, first, to control the correctness of deciphering brain activity signals and interpreting activity patterns when generating control commands, which can be controlled by the operator and, by correcting errors in the training mode, can improve algorithms of the intelligent system (self-learning classification system), and second, to control the operator's mental states recognized by the BCI intelligent system.

Finally, the general scheme of BCI as an intelligent system for real-time pattern recognition and classification with deep interaction (cyborgization) with cyborgized biological object (BCI operator) is shown in Figure~\ref{fig:BCI1}b. Closed-loop interaction of functional blocks of BCI control scheme and BCI operator leads to simultaneous self-training BCI operator for needed  brain states formation and learning BCI control intelligent algorithms as shown in Figure~\ref{fig:ni}.

\subsection{Active BCIs}

In the active BCI scheme described above, detection of sensorimotor rhythms during the imagination of motor acts is most often used to form control commands, which are easily detected in the motor cortex of the brain using ECoG/EEG/MEG~\citep{benabid2019exoskeleton, chholak2019visual}. 
Such BCIs are widely used to control external robotic devices --- manipulators, exoskeletons, wheelchairs since such imaginations are well associated with movements in space. Other controls can be specific brain activity patterns related to mental intentions, mental score, the imagination of music, etc., which can register using non-invasive fNIRS in the prefrontal cortex~\citep{naseer2015fnirs}. Such processes are more often used to implement communications, navigation in menus, Internet browsers, simple games, etc., for people with disabilities. An important requirement for the BCI control brain signals is the possibility of their use by the BCI operator for controlling neurophysiological processes underlying these signals. This can be achieved by special training, for example, using biological feedback. Therefore, the well-known slow cortical potentials (SCP), which are associated with changes in cortical activity related to the local excitation/inhibition in cortical neural populations, are used less often as control commands~\citep{Birbaumer1990slow}. The operation with the SCP-based BCIs requires prolonged continuous training. At the same time, it is known that such control techniques are well mastered by ALS patients with total motor paralysis~\citep{Kubler1999}.

\subsection{Reactive BCIs}

Unlike active BCIs, reactive BCIs for the formation of control commands detect and classify brain responses, the so-called event-related potentials (ERP), to external stimulation, usually visual, but sound, tactile, or other stimuli can also be used \citep{zander2011towards}. In this case, the functional scheme of the reactive BCI, inheriting many features of the active BCI, undergoes some change which provides the implementation of the stimulation system and the synchronous operation of other BCI blocks with presented stimuli \citep{hramov2021physical}. The visual evoked potentials (VEPs) and P300 evoked potentials (EPs) are most commonly used between ERP types for BCI operations. VEPs are brain activity responses that occur in the visual cortex after receiving a visual stimulus. Steady-state VEPs (SSVEPs), which are the brain's reaction to a high-frequency stimulation above 6 Hz, are most often used to create BCIs because the frequency components of SSVEPs remain almost constant in amplitude and phase over long periods of time. Using the SSVEPs, it is possible to develop BCI for the fast selection from a large number of possibilities, for example, for navigation in a menu when implementing various assistive technologies. %\citep{belkacem2020brain}. 
P300 evoked potential is an ERP component elicited in the process of decision making. P300 EP represents positive peaks in the EEG due to infrequent auditory, visual, or somatosensory stimuli. 
P300 EPs are usually elicited using the oddball paradigm, in which low probability target items are mixed with high probability non-target items. 
%\citep{Tamara2013human}. 
The main application of reactive BCI is communication with patients suffering from severe neurological diseases, which cause difficulties in communication with other people. The application for communication purposes usually implies a virtual keyboard on the screen, where the BCI operator selects a letter from the alphabet using an ERP-based neural interface. 
 
The advantage of SSVEP- and P300-based BCIs is that the BCI operator does not need long training because the ERPs are generated spontaneously. It should be noted that a key obstacle for BCI-based communication in humans is a low communication rate. A significant advance in fast communications was achieved by \citep{chen2015high}, who developed a noninvasive SSVEP-based BCI speller that allowed naturalistic high-speed communication with information transfer rates up to 5.32 bits/s. They developed a synchronous modulation and demodulation paradigm to implement the BCI speller based on the extremely high consistency of frequency and phase observed between visual flickering signals and the elicited single-trial SSVEP. Specifically, they proposed a new joint frequency-phase modulation method to tag 40 characters with 0.5-s long flickering signals and developed a user-specific target identification algorithm utilizing individual calibration data. A recent success achieved by using the  ERP-based registration systems allowed their commercial applications. %\footnote{See, e.g., ``Unicorn  Hybrid  Black''.  Available  online: https://www.unicorn-bi.com/ (accessed on 12 August 2021), and ``Neurochat''. Available online:  http://neurochat.pro/ (accessed on 12 August 2021).}. 

\subsection{BCI development}

Now, there are a lot of significant advances in BCI development. Moreover, the BCI field is one of few domains in which the delay time between a scientific idea, the first experiment, and application did not exceed several decades \citep{lebedev2017brain,hramov2021physical}.
%\todo{Well this is strong sentence, please add reference here}
However, many new BCI technologies are at the stage of laboratory experiments and are rarely implemented in real life or clinical practice. In order to fully realize the great potential of human-machine interaction using BCIs, the development of new BCI technologies in the forthcoming decades have to overcome a number of obstacles. Currently, significant progress has been achieved in deciphering the patterns of motor activity, recorded by invasive methods from a large number (from 100 to 4000 \citep{lebedev2017brain}) of electrodes. 
%\todo{What number?}
Consequently, the quality of signals and the possibility of precise control of external devices using implanted electrodes, even with reasonably simple classification algorithms, allow effective control of external devices, such as a prosthesis or an anthropomorphic manipulator. This became possible because the excitation of neurons in the motor cortex sufficiently determines the position, acceleration, and rotation angle of a primate's or human's limbs \citep{lebedev2017brain}. At the same time, the classic invasive interfaces do not allow the recording of neuronal activity and the electrical stimulation of the brain simultaneously. This problem has been overcome in optogenetic neurointerfaces, where the optogenetic methods are used for controlling neural activity, mainly in rodents \citep{grosenick2015closed}. However, only the first experiments on primates have been made \citep{yazdan2016large}, but this method has not yet been implemented for humans. 

At the same time, according to the recent review of \citep{delbeke2017and}, optogenetics predominantly used as a research tool in animals can be applied in humans in the nearest future.  On the one hand, this technique allows to control a certain group of neurons, and on the other hand, to simultaneously register the neuronal activity. Moreover, unlike electrical stimulation, optogenetic impact does not cause artifacts in recorded signals. This makes it possible to register changes in neuron dynamics caused by the stimulation with a minimal delay that allows controlling movements with high temporal resolution. Recently, an alternative neuromodulation technology that can be used in non-invasive interfaces is transcranial focused ultrasonic stimulation (tFUS) \citep{lee2017non}. Such the FUS-based technology is widely used to organize the information flow in brain-to-brain interfaces (B2BIs) due to high spatial accuracy, and non-invasiveness of neuromodulation \citep{nam2021direct}.

% \subsection*{Non-invasive BCIs}

A wide range of human-machine interaction applications is only possible for non-invasive neuronal interfaces that do not require neurosurgical intervention for implementation. However, since the accuracy in the command interpretation in non-invasive BCIs is currently much lower than in invasive ones, the application of non-invasive BCIs is limited to those areas where a large number of erroneous commands is uncritical. Highlighting near-term prospects for non-invasive BCI technologies, we can suppose that future developments will be devoted to those applications where neural signals provide us with information that is difficult or impossible to obtain using other methods (for example, instant alertness \citep{maksimenko2017visual} or fatigue \citep{dehais2018monitoring}), where perfect accuracy is not required for successful BCI operation. Apparently, the efficiency of such BCIs will be higher if their calibration is more accurate for an individual user, accounting for individual features of his/her brain activity. Therefore, increasing accuracy is a crucial problem for creating a perfect BCI. The brain states classification accuracy of about 80\% is the typical value for classifiers used in noninvasive BCIs. This accuracy is considered as acceptable, for example, in communications or neurorehabilitation, but is insufficient for controlling external devices (wheelchair and car with BCI control, or exoskeleton). Therefore, the important task is to increase the accuracy in the recognition of mental commands. 

\subsection{Multimodal BCIs}

One of the promising areas here could be the creation of multimodal BCIs, the so-called hybrid BCI, which would use several types of signals for command formation. 
Another more effective BCIs are multimodal systems that enable the estimation of the operator's mental state. 
Such a BCI is a hybrid of active and passive BCIs when the command classification algorithm of the active BCI is dynamically modified depending on the operator's state diagnosed by the passive BCI. 
Following \citep{choi2017systematic}, 59\% of created BCIs use only one type of physiological signals, mostly EEG. 
At the same time, one of the current trends of the BCI technology for improving the quality and effectiveness of the BCI is the combination of different approaches to create a hybrid BCI which takes advantages of different techniques. 

In this context, we would like to highlight three main approaches to increase effectiveness  of non-invasive hybrid BCIs: 
(i) BCIs using various signals that reflect brain activity, for example, EEG together with optical NIRS signals, or the use of two types of control processes, such as, e.g., modulation of sensorimotor rhythm together with SSVEP \citep{khan2014decoding}. 
(ii) BCIs using various physiological signals, for example signals of brain activity simultaneously with muscle activity (EMG) \citep{gordleeva2020real}.
(iii) BCIs using signals of brain activity in conjunction with external signals of different nature, such as, eye tracking, gyroscope, etc. Finally, for the several tasks we can create the brain states classifier based on recognition of functional brain network reconfiguration \citep{hramov2021functional}. This approach has proven itself well for controlling the epileptic brain \citep{maksimenko2017absence}, but we can expect this approach to be effective for cognitive states monitoring as well \citep{buch2018network}.

\subsection{Neurohybrid systems}\label{sec:Neurohybrid_systems}

Following non-invasive and invasive modern BCI technologies neurohybrid systems (NHS) represent the next level of integration between living systems and technical devices. They provide an interface at the cellular level between neurons and neuronal networks with artificial technical devices represented, for example, by electronic circuits. So that, the NHS is composed of two compartments that can communicate in bidirectional way to achieve a certain functionality.  

In technological applications, the artificial part of the NHS can be used simply to monitor the activity of neurons with cellular resolution in space and milliseconds resolution in time to resolve action potentials. Take, for example, multielectrode array sensors capable to monitor simultaneously thousands of electrophysiological channels. In turn, the living part of the NHS can get stimulation (driving) signals from the interface. It opens a possibility to control neuronal networks by external devises. All these technologies of recording and stimulation neuron networks in vitro are widely used in modern neuroscience. A spectacular example of the neurohybrid approach was the concept of neuroanimat proposed in 2001 by Steven Potter \citep{demarse2001neurally}. The collected neuronal signal was further sent to artificial body, e.g. embodiment, representing a roving robot exploring the environments. Collecting signal from sensors external  artificial computational system generate a feedback pattern stimulating the neuronal culture to achieve adaptive result.  

In the late 90th there were several attempts to create a direct interface between living neuron and silicon electronics \citep{fromherz1996towards}. Key obstacle to do this was principle difference in mechanisms of electrical pulse generation in neurons and in electronic generators. Ionic currents responsible for action potentials in neurons are much slow than electrons/holes in transistors. However, possibilities of interaction through electrical field still obviously exist. In particular, electrical discharges in neurons can drive the gates of the FET transistors activating them. In such a way, an active silicon substrate can be created recording neuronal activity and stimulating it. One possible functional role of such co-functioning of neuronal and silicon compartments was to enhance the information processing efficiency of (i) living neurons by electronics and (ii) artificial neuronal networks by living neuronal circuits \citep{morozova2013stimulation}. 

The design of adaptive bidirectional interface between neuronal and electronic networks can give further insights in searching ``strong'' (e.g. living-like) artificial intelligence solutions. Neuronal network adapt itself following synaptic plasticity (for, example, STDP) changes. At the same time it can serve as a teacher for the artificial part of NHS tuning the electronic compartment to achieve a certain functions. Such experiments may shed the light on how the electronics should be tuned to follow the living cell functionalities. In modern state-of-art paradigm the design of memristive devices, specifically, crossbar of memristors capable to imitate neurons and synapses in more biologically relevant way definitely will give the novel technological push in the design truly adaptive NHSs.      

Furthermore, technologies of neurohybrid systems can be projected to in vivo studies. Taking a signal from the input of a brain circuit one can train artificial neuronal network and send the appropriate signal to the output. Theoretically, such technology will permit to enhance brain functionality by means of external artificial devices \citep{mishchenko2018optoelectronic}.

\section{Where is the revolution?}\label{sec:revolution}

\begin{figure}[ht]
    \centering
    \includegraphics[width=1.0\textwidth]{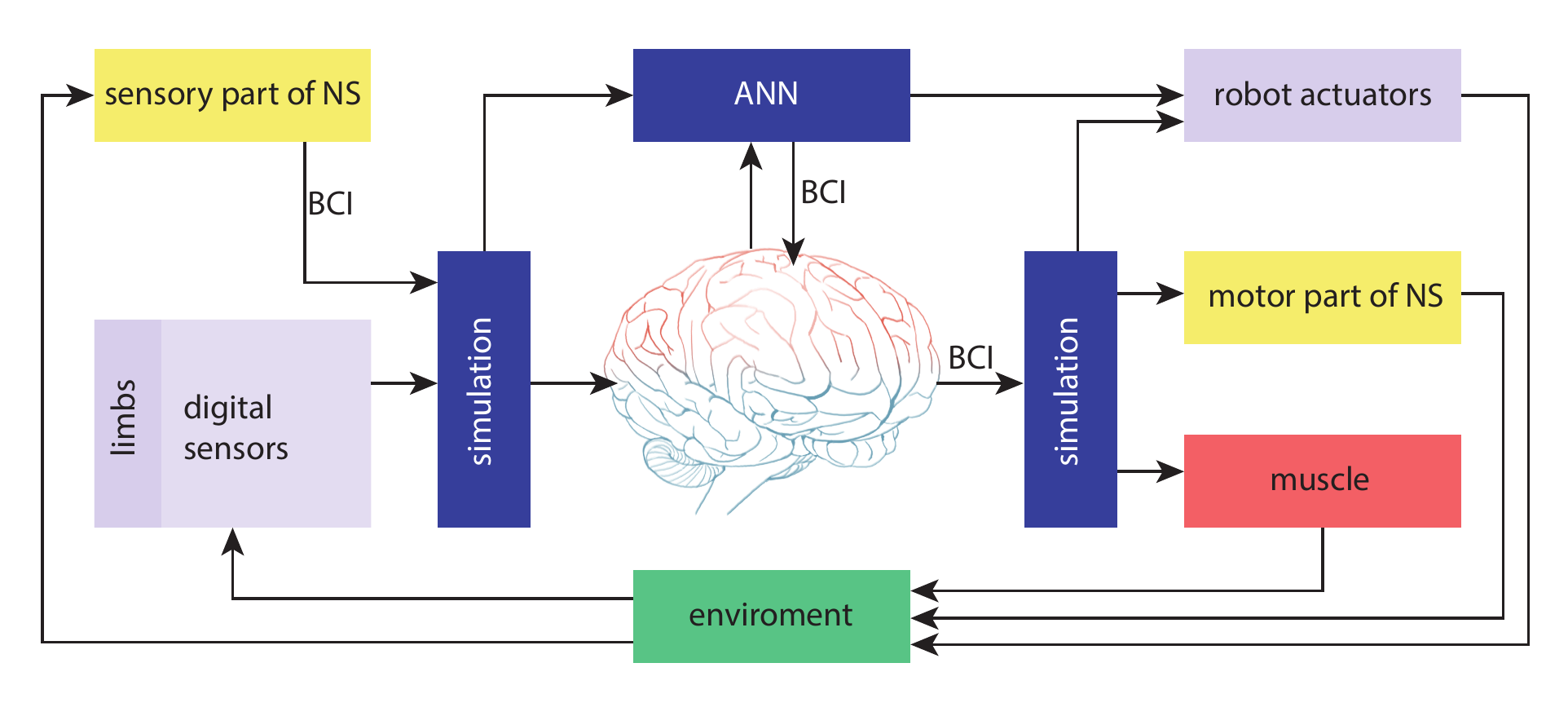}
    \caption{The technological map indicates the real-time neurosimulation as middleware. Blue is the simulation, yellow -- parts of the nervous system, red -- muscles, green -- environment, lilac -- digital parts of the system.
    Digital sensors mounted on limbs as well as sensory neurons are used as inputs for the simulation. 
    The artificial neural network (ANN) could be used for spike sorting. The BCI reads the neuronal activity and later transfers it to the simulation which generates the neuronal patterns to manage/stimulate via motor NS or/and muscles or robot actuators. 
    }
    \label{fig:layered_technologies_stack}
\end{figure}

\begin{figure}[ht]
    \centering
    \includegraphics[width=0.9\textwidth]{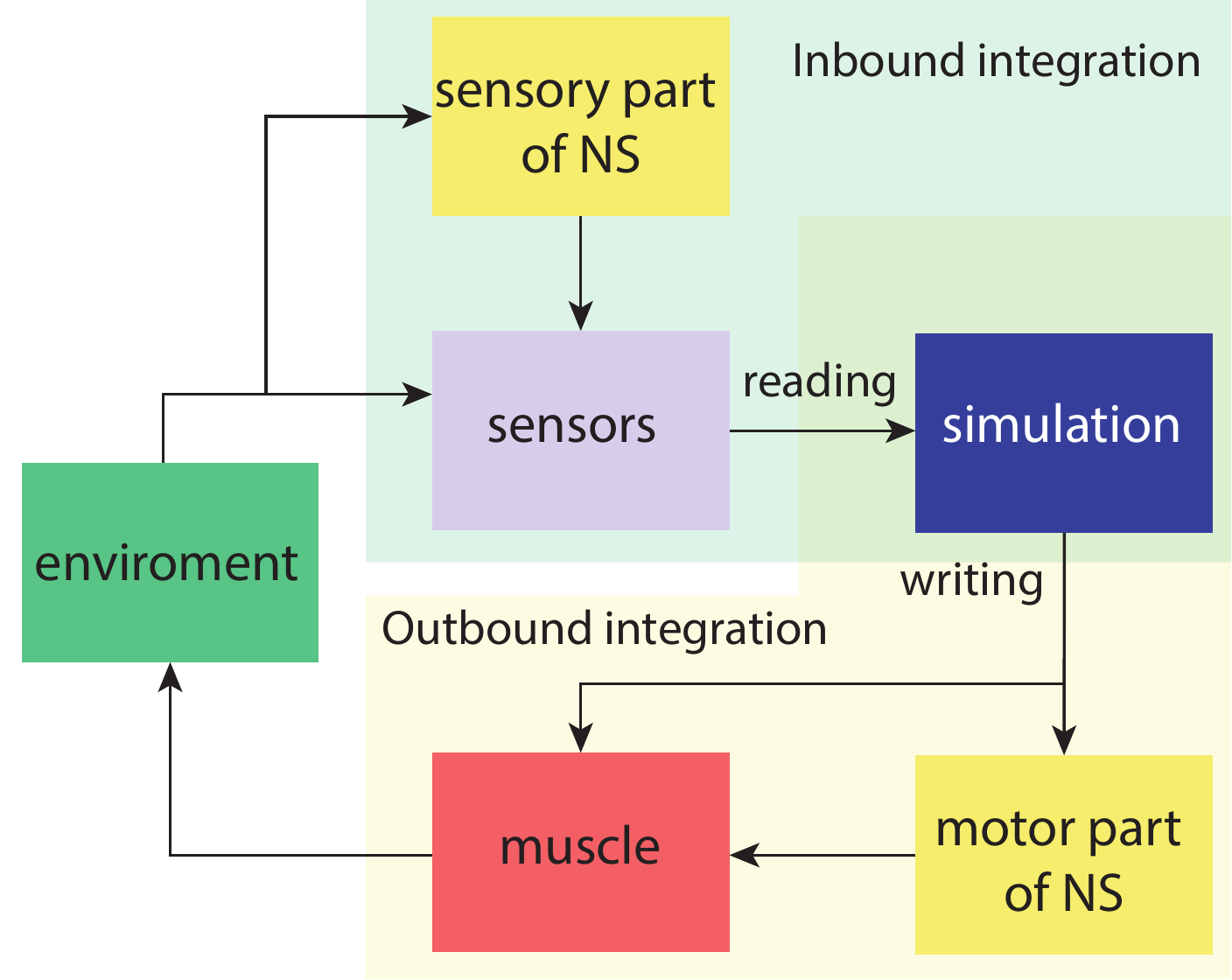}
    \caption{
    Typical closed loop bio object - simulation integrated system. Blue is the simulation model, yellow -- parts of the nervous system, red -- muscles, green -- environment. 
    \textbf{Env} stands for an environment, 
    \textbf{NS} -- a nervous system, 
    \textbf{Mus} -- muscles. 
    \textbf{Inbound integration} (green rectangle) consist of sensory part of NS, digital sensors and simulation;
    \textbf{outbound integration} (yellow polygon) -- real-time simulation, motor part of NS or/and muscle (possibly robotic actuator).}
    \label{fig:ni}
\end{figure}

%quotation fixed
Before we can demonstrate that our project is a true technological revolution, we should define first the exact meaning of such concept. According to \citep{perez2010technological}, page 26: ``Thus, on a first approximation a technological revolution (TR) can  be  defined  as a set of interrelated radical breakthroughs, forming a major constellation of interdependent technologies; a cluster of clusters or a system of systems.''
Therefore, the revolution we propose is in the shift of the perspective we use to look at bio-plausible neurosimulations and mathematical models of biological processes that could act not only as some abstraction and demonstration of the complex processes for research purposes, but because, at the same time, our project is innovative in the real-time implementation of such mechanisms being integrated as a part of a biological system (Fig.~\ref{fig:layered_technologies_stack}).
The aggregation of several technologies, like the fourth technological revolution about BCI, neurostimulation implemented in spiking neuromorphic computing technologies, neuroprosthetics, and the electrical nerve and muscle stimulation (EMS), creates the new phenomenon that we call the ``neuropunk revolution''. 
Solving the problem of integration of multiple electrodes in brain and their bio-compatibility could open clear perspective for next level BCI. 
Besides the digital implementation via real-time simulation using mathematical models we see the bio-inspired spiking technologies like memristive, FPGA and ASIC schematics as the possible low-level implementation of simulated neuronal circuits to be integrated into a biological tissue via BCI or neurostimulation/EMS. 
An electronic device reproducing  functions of a biological system (external or implanted in biological objects) via the real-time simulation, 
has a cybernetic nature but ``speaks  the same language'' with the nervous or other biological systems. 
In this case the simulation device could extend the capacities of a biological tissue, damaged or even the healthy one. 
This brings to life \emph{new processes} for the neurorehabilitation, post trauma rehabilitation, sport training, or integration with robotic systems.
This approach is new and currently not implemented.

We suppose the new concept of integration of real-time neurosimulations into biological systems could play the role of a middleware integrating the digital world's: sensors, actuators, ANNs with a biological world's: sensory, motor neurons and muscles (Fig.~\ref{fig:layered_technologies_stack}).

A BCI could integrate a neuronal activity into a neurosimulation which in its turn integrates in robotic actuators via digital channels or into motor neuron pools or muscles via stimulators. The ANN plays the role of a filter and a spike sorter that could be integrated with a real-time neurosimulation and/or brain via BCI devices or robotic actuators.

\subsection{Closed loop system}\label{sec:closed_loop_system}

The possible close loop system with a real-time neurosimulation of a biological object is presented in Figure~\ref{fig:ni}.
The simulation model receives (reads) signals from sensors that indicate changes in the environment or/and in the sensory neurons, for instance, pain or cutaneous input.
Neural activity generated on sensory data is later transmitted (written) into neuronal motor pools or directly to the muscle.
The motor response affects the environment thus closing the loop.

\subsection{Outbound integration}

The outbound integration of real-time simulation into a biological tissue (nervous system or muscles) (Fig.~\ref{fig:ni}~outbound integration)
provides the option to compensate a missing part of neuronal circuits with simulating mathematical models.
As a possible application we remark the case of spinal cord injury (SCI), when a real-time simulation of the spinal cord segment could generate motor patterns to control muscles below the trauma.
In a similar way, we could treat a cerebral palsy or a foot drop compensating the missing or injured parts of the nervous system. 
In the case of the stroke real-time simulations of the spinal cord and cortical columns, models could restore the missed control over limbs and other body parts.

\subsection{Inbound integration}

The spike sorting and the pattern recognition could be done ``naturally'' by means of neural circuitry that could be used as an outbound interface to digital systems (Fig.~\ref{fig:ni}), for example robotic \citep{mikhaylov_neurohybrid_2020}.
The pattern recognition circuits  could generate as well managing patterns with help of neuronal generators~\citep{talanov_oscillator_2020} and could manage actuators of an exoskeleton or a  robotic arm via spiking to digital converter. 
The other way around, the pattern could be used for analysis via traditional ANN programs.

\subsection{In-outbound integration}

The illustrative example of the in-and-outbound integration is the medical use in case of SCI.
The brain activity is detected and processed via the brain implant~\citep{lebedev2017brain} and
later transmitted via Bluetooth into the digital/spiking electronics simulation of the spinal cord segment where the locomotion pattern is generated according to the input from the brain, gyroscope or/and afferent (pressure sensors, and goniometers attached to foot, ankle, knee, and hip).
The generated pattern is used to stimulate limbs muscles to reproduce and, thus, restore the locomotion in the walking pattern.

\section{Unconventional AI}\label{sec:ai}

\subsection{Spatial biocomputing}

We see one of the newest and promising approach in artificial intelligence and unconventional computing is Physarum computing. It is based on a reaction-diffusion computer, which is a spatially extended chemical system, processing information using interacting growing patterns, excitation
and diffusive waves, for example, Belousov–Zhabotinsky reaction \citep{adamatzky2005reaction}. 
In reaction-diffusion processors, both data and results of the computation are encoded as concentration profiles of reagents. 
The computation is performed via the spreading and interaction of wave fronts. A great number of chemical
laboratory prototypes, designed by De Lacy Costello, are discussed in~\citep{Adam_Physarum}. \\
Andrew Adamatzky \citep{Adam_Physarum} indicated several key differences between the reaction-diffusion computer or even unconventional and conventional computers:
\begin{enumerate}
    \item \emph{parallelism}: there are thousands of elementary processing
units, micro-volumes, in a standard chemical vessel;
    \item \emph{load connections}: micro-volumes of a non-stirred chemical medium
change their states, due to diffusion and reaction, depending on
states of, or concentrations of reactants in, their closest neighbors;
    \item \emph{parallel I/O}: in chemical reactions with colored product the results of the computation can be recorded optically;
    \item \emph{fault tolerance}: being in liquid phase, chemical reaction-diffusion computers restore their architecture even after a substantial part of the medium is removed; \\
\end{enumerate}
Andrew Adamatzky showed in this book~\citep{Adam_Physarum} possible usage of Physarum (slime mold) as reaction-diffusion computer. Physarum has several stages of its life-cycle. One of the first is plasmodium (a ``vegetative'' phase), which is a single cell with a myriad of diploid nuclei. The plasmodium looks like an amorphous yellowish mass with networks of protoplasmic tubes. 
He showed possibility to control plasmodium mobility by propagating special waves. They put repellents - source of the light and attractants - oat flake on the other edge. The plasmodium distant from the source of light. \\
When the plasmodium is placed in an environment with distributed sources of nutrients, it forms a network of protoplasmic tubes connecting the food sources. 
If we interpret the sources of food (e.g. oat akes) as nodes and protoplasmic tubes as edges, we see that the plasmodium constructs a planar graph on the sources of food. Adamatzky stated in his research that the topology of the plasmodium's protoplasmic network optimizes the plasmodium's harvesting on the scattered sources of nutrients and makes more efficient the flow and transport of intracellular components. \\
At the moment Physarum computers can: (1) find shortest path~\citep{Adam_Physarum,adamatzky2005reaction}; (2) implement logical gates~\citep{Adam_Physarum,vallverdu2018}; (3) construct spanning trees~\citep{Adam_Physarum}; (4) implement primitive memory~\citep{Adam_Physarum,vallverdu2018}; (5) implement spatial logic and process algebra~\citep{Adam_Physarum}.

\subsubsection{Emotion machine}

The book of Marvin Minsky ``The emotion machine'' \citep{minsky2007} is full of interesting ideas regarding artificial intelligence, artificial cognition and their connection to emotions. 

First of all we want to emphasize the $Critic \rightarrow Selector \rightarrow Way To Think$ triplet, of mental processes, where critic is main analytical mechanism of a brain that process all inbound data of an organism, selector is the mechanism that allocates brain resources in simplest understanding neurons that could play the roles of other critic or way to think, way to think is the mechanism to update the stored information. 
In everyday life individuals face problems and tasks, such as ``Open the door'', ``Eat with a spoon''. Critic is a reaction to this problem. For example in the software maintenance domain, when an auto generated incident is processed the Auto Generated Incident critic is activated. 
After the activation, a critic becomes a selector. 
A selector is capable of retrieving resources (a critic or a way to think).
The Way To Think according to Minsky is a way how a human thinks. For example, ``If I know the problem, I use my experience in analogy situation to solve it''. 
The triplet is operating in the boundaries of the ``Model of 6'' or six levels model of mental processes: (1) Instinctive reactions; (2) Learned reactions; (3) Deliberative thinking; (4) Reflective thinking; (5) Self-Reflective thinking; (6) Self-Conscious reflections.

The level 1 integrates all sensory inputs and processes them according to inborn algorithms and later transfers them to the learned level. 
The learned reactions level is used for all the learned activities. The deliberative level uses the learned skills and information. In this level we do the logical reasoning and when it is focused on the events of nearest past we use reflective thinking level. 
The self-reflections places us as the subject in the reflective thinking centre and there is no strict border line between reflective thinking and self-reflective thinking levels. 
The highest level is dedicated to self-conscious reflections and mainly concentrated on psycho-emotional states regarding the high level goals, ideals, values and censors.

Previously we used this approach in Software maintenance automation domain~\citep{tu-link-gen-2}. 
The inbound textual information is processed to build the machine understanding and for this we used several types of way-to-think: (1) Simulation; (2) Correlation; (3) Reformulation; (4) Thinking by analogy.

\emph{Practical example 1}. If an incident is automatically generated, the system should process it using instruction book A. 

\emph{Practical example 2}. If a system recognizes the problem, use analogy to solve it. 

\subsection{Fungal Computing}
To further comprehend the essence of unconventional AI, we first explain the concept of agents, which are the entities that make decisions and are the anchoring force behind putting AI into action.

Intelligent agents, according to conventional AI, can learn while performing tasks with some level of autonomy, allowing them to do specific, predictable, and repetitive tasks~\citep{russell2020artificial}. 
Agents in unconventional AI, on the other hand, have the property of `naturally occurring' learning to adapt to environmental changes in order to survive. 
This evolutionary property strengthens their autonomy so that they can not only sense the environment/user and react reactively at times but also perform specific activities, initiating the integration of an actuator~\citep{adamatzky2021reactive}. 
Apart from these two significant differences between conventional and unconventional agents, four other characteristics of an intelligent agent can be considered common in these two categories which are:
\begin{enumerate}
    \item They can interact with other entities involved in the environment.
    \item They can accommodate new rules incrementally.
    \item They possess goal-oriented behaviors.
    \item They are knowledge-based to communicate, process, and act on this knowledge.
\end{enumerate}

In addition to these characteristics, unconventional intelligent agents demonstrated fault tolerance~\citep{adamatzky2021adaptive}. In fact, they can rebuild their architecture and intelligence even after a significant portion of the medium has been removed or damaged. This feature is essential for organic electronics employed in applications and the development of unconventional living architecture, soft and self-growing robots, and intelligent materials derived from fungi or bacteria.

In the FUNGAR project, Adamatzky et al. showed that fungi exhibit features similar to memristors (memory resistors)~\citep{beasley2021electrical}, electronic oscillators~\citep{dehshibi2021electrical}, pressure~\citep{adamatzky2021living}, optical and chemical sensors~\citep{dehshibi2021stimulating}, and electrical analogue computers~\citep{adamatzky2021fungalelectronics}. Fungal electronics can be used as standalone sensing and computing devices or integrated into fungal materials and wearables~\citep{adamatzky2021reactive}. They are a promising candidate for producing sustainable textiles and eco-friendly bio-wearables attributed to these characteristics. Wearable is composed of or including fungi-colonized fabric could function as a large distributed sensorial network, providing a great potential to incorporate natural systems' parallel sensing and information processing capabilities into future and emerging intelligent agents~\citep{adamatzky2021reactive,adamatzky2021fungalelectronics}. The FUNGAR research team has also shown that reactive fungal wearables are capable of low-level perception and have the potential to be developed into autonomous adaptive device~\citep{adamatzky2021reactive}. This feature would allow fungal wearables to communicate with conventional electronics. In terms of extension and interconnection, fungal networks are certainly a sustainable infrastructure-forming substrate, capable of wire loci separated by considerable space. 

Adamatzky et al. have also demonstrated that mycelium networks not only sense but also process information. 
They observed that the \textit{Pleurotus djamor} oyster fungi produce action potentials similar to electrical potential impulses~\citep{dehshibi2021electrical}, i.e., spontaneous spike trains with high (2.6 min period) and low-frequency components (14 min period). 
This evidence suggests that it is possible to turn fungal reactions into Boolean circuits~\citep{beasley2021electrical}, allowing fungal agents to function as parallel biological processing networks~\citep{adamatzky2021fungalelectronics}. 
However, the lack of an algorithmic framework for the extensive characterisation of the electrical activity of the substrate colonised by mycelium of the oyster fungi \textit{Pleurotus djamor} inspired them to develop a framework to extract spike patterns, quantify the diversity of spike events, and measure the complexity of fungal electrical communication in order to build an experimental prototype of fungi-based information processing devices~\citep{dehshibi2021electrical}. 

According to their findings, the complexity of the Kolmogorov fungal spike spans from $11 \times 10^{-4}$ to $57 \times 10^{-4}$. In~\citep{dehshibi2021stimulating}, the Kolmogorov complexity of the human brain is measured in normal, pre-ictal, and ictal phases, yielding 6.01, 5.59, and 7.12 values, respectively. Although the complexity of fungi is much lower than that of the human brain, variations in the mycelium sub-network show a degree of intra-communication~\citep{dehshibi2021electrical}. In fact, different parts of the substrate transmit different information to different parts of the mycelium network, where prolonged excitation wave propagation leads to higher levels of complexity. In their experiments, they compared the fungal language to European languages in terms of Lempel-Ziv and Kolmogorov complexity. They speculated that the complexity of fungal language is greater than that of human languages (at least in European languages)~\citep{dehshibi2021electrical}.

In another study~\citep{dehshibi2021stimulating}, they explored the communication protocols of fungi as reporters of human chemical secretions like hormones to see if fungi can sense human signals. The researchers exposed \textit{Pleurotus oyster} fungi to hydrocortisone, which was applied directly to the surface of a fungal-colonised hemp shavings substrate, and measured the fungi's electrical activity. Hydrocortisone is a hormone replacement medication that works similarly to the natural stress hormone cortisol. The response of fungi to hydrocortisone was also explored further using X-ray to identify changes in the fungi tissue, where the substrate receiving hydrocortisone can inhibit the flow of calcium and, as a result, reduce its physiological changes. The computed tomography revealed that hydrocortisone exposure caused a change in grey tones, which is linked with active ion transport triggered in living mycelium in response to the hormone. Simultaneously, they could identify consistent metrics for tracing the stimulus and tracking the propagation of ionic waves in response to the hormone by recording biopotential signals and extracting their numerical complexity.

\subsection{Emotions, affects and NEUCOGAR}

Despite of common view emotion and affects play important role in the decision making process. Especially interesting from our perspective is the work of \citep{emotionsbraintorobot}, where authors discuss the role of neuromodulators: dopamine (DA), serotonin (5-HT) and their impact over emotions and mammalian brain and as a result decision making process.

The cognitive architecture NEUCOGAR \citep{bica2015neucogar} was created to implement psycho-emotional states as much bio-plausible as we can taking in account current state of technology. 
We used the ``cube of emotions'' created by Hugo L{\"o}vheim \citep{lovheim2012} based on a three-dimensional model of emotions and levels of monoamine neuromodulators (serotonin, dopamine, nor-adrenaline (NA)). Vertexes of the model are basic affects, defined in Tomkins theory of affects \citep{primer_affect_psychology}. 
According to his theory, these are the eight basic emotions: enjoyment/joy, interest/excitement, surprise, anger/rage, disgust, distress/anguish, fear/terror and shame/humiliation \citep{primer_affect_psychology}. 
L{\"o}vheim gives extended explanation of mapping of each neurotransmitter to a group of affects.
We assume the indirect influence of the neuromodulation and thus neuromodulators over the computational processes of the host computational system where the bio-plausible network with main monoamine neuromodulators systems are implemented in the following way:

\textbf{Dopamine}. 
Dopamine plays a major role in motor activation, appetite motivation, reward processing and cellular plasticity, and important in emotion. 

According to the L{\"o}vheim ``cube of emotions'' dopamine is associated with ``[r]eward, reinforcement, motivation''. From a computational system perspective dopamine plays a role in: reward processing thus in decision making, working memory - memory distribution and storage in computing system, motivation - decision making.

\textbf{Serotonin}. ``Serotonin has been implicated in behavioral state regulation and arousal, motor pattern generation, sleep, learning and plasticity, food intake, mood and social behavior''. 
Serotonin plays a crucial role in the modulation of aggression and in agnostic social interactions in many animals. 

L{\"o}vheim associates serotonin with ``[s]elf confidence, inner strength, and satisfaction''. 
Thus serotonin influence the host computing system in the following way: decision making, confidence and satisfaction, this way serotonin impacts learning and storage of the information.

\textbf{Nor-adrenaline}.  According to L{\"o}vheim cube of emotions: ``Attention, vigilance, activity \ldots while nor-adrenaline has been coupled to the fight or flight response and to stress and anxiety, and appears to represent an axis of activation, vigilance and attention''. 

\begin{figure}[ht!]
    \centering
    \includegraphics[width=\columnwidth]{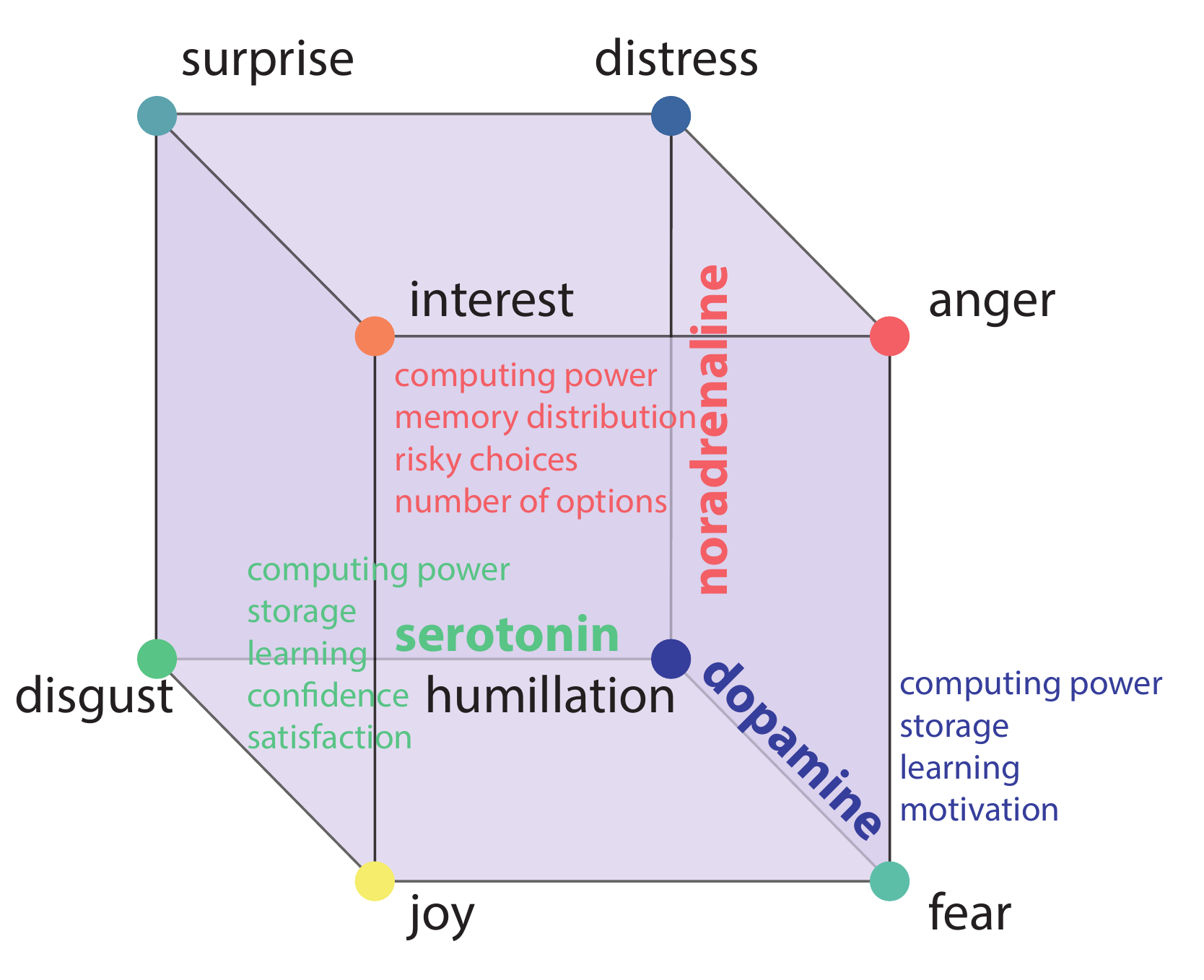}
    \caption{The mapping of computing system parameters to basic emotions and monoamine neuromodulators levels.}
    \label{fig:cube}
\end{figure}

The triplet neuromodulators $\rightarrow$ psycho-emotional states/affects $\rightarrow$ drives for example survival or reproduction, that heavily influence the decision making plays crucial role in everyday life of biological system of a mammal. 
We proposed the cognitive architecture NEUCOGAR \citep{bica2015neucogar} to compensate usual to conventional AI lack of emotional mechanisms. It uses the mapping of computer parameters to affects and levels of neuromodulators in the way represented in Figure \ref{fig:cube}. 

\section*{Integration into robotic devices}
\label{sec:robots}

The neuropunk revolution will have a significant influence on robotic devices covering a broad range of applications that includes basic BMI/BCI-based control schemes of teleoperation, wearable sensory or/and assisting devices, and their invasive versions. 
While for the aforementioned applications we expect impressive improvements of existing technologies and new technologies emergence, the most revolutionary breakthrough is expected in neurocontrol of autonomous robots. 
The term ``neurocontrol'' has a number of definitions, and in this section we refer it as the ability of controlling a robotic device using a human nervous system or its reasonably simplified model. BCI-based control schemes were previously discussed in Section BCI, machine to nervous system interface and the following subsections discuss the power of invasive spinal cord devices (with sensory and locomotion-assisting functionality) and a glance into the future of autonomous robots’ neurocontrol and emotional human-robot interaction.

\subsection{Spinal cord modularization}

The ideas of a spinal cord modularization are quite natural for the mammalian nervous system, where we usually view a spinal cord as an intermediate but separate from a brain system. The mammalian spinal cord plays an important  role in motor control and perception: it is not just a signal transmitting sub-system but it has a role of a primary processing center for an inbound sensory information and generates complex motor patterns. More than that, it is a self- learning system due to the spike-timing dependent plasticity (STDP). 

A large number of varying CPGs topologies could be pre-generated in an offline mode and integrated into a robotic spinal cord control system, which should allow an appropriate function for a particular task CPG selection and its adaptation in real-time.
Pre-generated CPGs, which are popular in bipedal locomotion systems~\citep{dzeladini2018cpg} %, duysens2018walking,santos2017biped} 
and range from simple walking primitives approach~\citep{khusainov20163d} to human CPGs learning by imitation~\citep{masumori2021personogenesis} and adaptive gait planning with hierarchical CPGs~\citep{yao2021humanoid}, would further require an integration with ``on-board sensors''~\citep{li2017neural}, a precise initial calibration~\citep{khusainov2017humanoid} for a particular user after an invasive installation and capabilities of self-calibration~\citep{stepanova2019robot}, adaptation~\citep{thor2019fast} and autonomous evolution in a long-term life cycle. 
An integration with a ``brain computing system'' of a robotic device via a neurointerface could be done using ``impasse'' states recognition: if the neuromorphic robotic spinal cord is unable to process a current sensory and/or motor state, an exception should be redirected to a central processing machine or the ``brain''.

\subsection{Autonomous robots neurocontrol}
\label{ssec: Neurocontrol }

Recent neurocontrol applications in robotics are mainly concentrated on constructing a teleoperational communication link between a human and a robot via BCIs/BMIs or applying neural networks as a part of a control system. 
While at the dawn of neural networks technology simple models with just few layers and neurons were treated as a breakthrough~\citep{crick1989recent}, nowadays it turned into a well-established technology with applications in recognition~\citep{li2020quantum}, path planning~\citep{sung2021training}, motion control~\citep{su2020improved} and other typical activities and functions of  a robot in various environments and settings~\citep{magid2021pandemic}. 
Yet, recent technologies could hardly replicate a power of a human neural system that could be feasible enough for integration into a mobile robot onboard control system and therefore      a number of heavy simplifications are to be implemented.

A robot definition by ISO 2012 and ISO 2021 emphasizes its level of autonomy, which is ``an ability to perform intended tasks based on current state and sensing, without human intervention''~\citep{iso2012iso}. 
Adapting Sheridan’s levels of automation of decision and action selection~\citep{parasuraman2000model} to robotics, we place teleoperation on the lowest level and a fully autonomous robot on the highest level of the scale. 
In between are attempts for partial automation of different intensity and purposes, from filtering out a noise of human control in master-slave settings~\citep{kikuuwe2015phase} and autonomous return upon losing a connection with an operator~\citep{alishev2018network} to autonomous exploration of the environment with a minimal human   disruption~\citep{selin2019efficient}. 
At the teleoperation level neurocontrol is performed entirely via a BMI, while at higher levels more sophisticated approaches are required.

Mobile robot control originated from a sense-think-act %~\citep{siegel2003sense} 
and reactive %~\citep{ranganathan2003reactive} 
paradigms, which by 1990s were unified into a sustained scheme of a hybrid approach %~\citep{huber2000hybrid}.
~\citep{siegel2003sense}.
By mid-90s, probabilistic robotics approach %~\citep{thrun2002probabilistic} 
significantly challenged the classical methods, which in turn is recently overpowered by learning-based and deep learning techniques. % [289]. 
While the later approaches were inspired by neural system pattern simplifications, we believe that the revolutionary future of robotics in terms of neurocontrol lies in creation of autonomous robots that will be operated with an in vivo human nervous system (or its reasonably simplified digital models of varying complexity levels), which is integrated directly into an on-board control system. Such systems will be capable to successfully trigger appropriate sense-think-act, reactive and learning-based techniques on-demand in real-time and allow other, yet unemployed in robotics, capabilities of a human brain and nervous system.

Interesting approaches that attempted to integrate a human-like pain sub-system into a robot control system were discussed in~\citep{Johannes2017}. By the analogy with a biological system, an implementation of a self-preservation instinct and   an alarm system of a body could play important role in a robotic system influencing not only immediate reactive behavior but also learning and information retrieval processes~\citep{cheng2020neuroengineering}. It also could have a role acting as a mirror attachment system that was able to connect humans suffering pain to their robotic counter-parts, being able to share and diminish the intensity of such reaction, though the exchange process of a shared experience~\citep{ishihara2011realistic}. The pain related coloring of memories and learning rate enhancement as well as coloring influence of information retrieval for the behavioral formation should influence the behavioral patterns of a robot implementing self-preservation instinct making the robotic system more enduring and stable.

\subsection{Honest human-robot interaction}
\label{ssec:HRI}

Another important and promising functionality of a human neural system inspired approach is to generate plausible emotion states of robots that interact with humans. 
Recently robots have started communicating via verbal and non-verbal interfaces, which included facial expressions, non-verbal behavior and semiotic indicators~\citep{song2017expressing}. 
Yet, even when researchers employ a combination of  philosophy, art and science approaches~\citep{chesher2021robotic}, resulting facial expressions as well as other means of human-robot communication could only demonstrate predefined emotional reactions. These reactions are subjective, gender and culture-dependent and are rarely confirmed with statistically valid number of experiments with real users. Moreover, attempts to coordinate different modalities of communication to achieve an expressive manner of human-robot communication that could resemble or replicate a human-human communication (e.g., reported in~\citep{aly2020designing}) are too complicated and are doomed to fall behind expressive human-human interactions. We believe that generated in real-time via a human neural system based control system reactions (e.g., proposed in~\citep{Emotico}) would significantly outperform any prerecorded reactions and demonstrate a different level of honesty, empathy and understanding between a human and a robot, which are critical for comfortable daily interactions with robots.

\section{Challenges: neuroimplants \& BCI} \label{sec:neurorehabilitation}

Over the past decades, electrical stimulation has shown a range of new possibilities, including protocols to control pain syndromes and activate neural circuits to restore motor and vegetative functions after injuries. With the spread of neuromodulation methods and their translation into the clinic, it is within reach of patients, while understanding the mechanisms of neurostimulation is still limited. A large number of technological problems still need to be solved to create  effective, quickly achievable and safe neuromodulation techniques. In addition to reliable and cheap devices and electrodes in the near future, it will be important to functionally integrate neurostimulators with feedback systems (such   as the integration of sensory-proprioceptive inputs) to improve motor functions, balance, coordination, and sensorimotor integration. Some of these approaches were gradually implemented in rehabilitation to restore and optimize sensory and motor control, which is carried out through conducting pathways and neural systems. At the end, the universal control systems with effective integration with interfaces will be able to restore and/or provide prosthetics of sensory and motor functions after various CNS disorders. Finally, the successful clinical translation of these methods not only allows the development of novel neuromodulation methods, but also raises new ethical problems that must be taken  into account for the optimal restoration of both neurological functions and the  quality of life of patients. At the end of this prospective review, we summarize  the main problems and challenges in development and further implementation of  new devices within the concepts of neuroimplants, neuromodulation, and BCI. Traditionally, electrical stimulation is applied through the skin, in places close to the location of peripheral nerves, with intramuscular electrodes for direct muscle stimulation, or directly to the different part of peripheral or central nervous system with invasive electrodes. 
The standard approaches of peripheral stimulation have several limitations, which include: (1)~development of rapid muscle fatigue, (2)~nonlinear motor response and sub-optimal control of   stimulation, (3)~poor individual muscle control selectivity and limited levels of freedom during complex movements, and (4)~technological limitations such as external control mechanisms and bulky external devices and power supplies. 
These disadvantages limit the widespread and clinical use of these devices. In the recent years, new paradigms of invasive and noninvasive stimulation have shown promise in the restoration of motor functions, activating the operation of complex sensorimotor networks, for example in the spinal cord, facilitating the central pattern generators (CPGs). 
Differences in neural structure involved in the effect of neuromodulation and specific paradigms between central and peripheral stimulation, ultimately leading to more sustained motor activity with central stimulation. 
Paralysis after spinal cord injury (SCI) remains a critical condition that has a devastating effect on quality of life, independence, and physical and psychological well-being. 
Recent evidence suggests that neuroprostheses can significantly support the potential to restore neurological functions after SCI, including upper and lower limb movement, body weight load, ability to walk, autonomous  control of breathing, coughing, and bladder and intestine. 
Available devices   already successfully provide independence from artificial ventilation in people with high tetraplegia and respiratory paralysis. 
Modern neuromodulation approaches can improve standing and even restore individual motor functions of the lower limbs. 
The stimulation of sacred segments is successfully used in people with a neurogenic bladder in order to arbitrarily start urination. Current  technologies, however, are still far from effectively replacing the wheelchair as the primary vehicle for daily mobility in patients with SCI~\citep{Guevremont_2006}. 
Until now, relatively simple neurostimulation approaches, such as  stimulation for diaphragm control, have been successfully performed through minimally invasive implantation. 
In contrast, motor control requires coordination of complex synergistic patterns and complex neuromuscular interactions. 
The CNS stimulation has shown great promise in restoring neurological functions, and may ultimately provide significant functional recovery after SCI.

Several factors currently affect the absence of widespread clinical use of neuroimplants: (1)~implantation of devices is associated with an initial high surgical cost that requires long-term depreciation; (2)~implantation may require repetition of invasive surgery to replace the battery; (3)~neuroprosthesis usually requires a long, sometimes ineffective or inconvenient phase of stimulation optimization before users can ultimately benefit clinically; (4)~insufficient understanding of the mechanisms of stimulation~\citep{Gater_2011}. 
It is also necessary to solve the problems associated with the successful commercialization of neuromodulation devices for wide distribution in clinics. 
At the end, existing devices should achieve significant functional improvements and ensure efficiency, security, and accessibility, as well as be convenient for the user in daily life with minimal maintenance. 
The development of neuronal prosthetics is technologically complex and requires multidisciplinary efforts with several important conditions that take into account the functional aspects of the nervous system, the peculiarities of modern technologies, aspects of stimulation, power supply, signaling, surgical equipment, as well as safety issues. 
Compared to implants for brain stimulation, spinal cord neuroimplants require a more sophisticated fabrication technique due to the high mobility of the spinal cord and spine. 
Achieving an optimal degree of motor control or relieving chronic pain is the cornerstone of any neuronal prosthetics system. 
With respect to motor function, this control must be defined in relation to coordination and motion vector controlled in all degrees of freedom and force, and the volume of muscle contraction must be resistant to fatigue.
The technological considerations for stimulating devices are ultimately determined by the necessary tasks. 
The neurostimulation effect is usually correlated with the location, volume and geometry of the electric current applied to the neural structure. 
Therefore, universal and highly efficient stimulation requires flexible interfaces with adaptive control systems. 
The development of novel neuroprostheses begins with technical reviews of the corresponding stimulation approaches, capable of providing control with sufficient number of stimulation channels, safety, possibility of detecting system errors, and clinical effectiveness. 
In addition, the device will need to be capable of flexible stimulus generation by applying optimal forms of stimulus appropriate to the intended clinical purpose. 
In addition, neurostimulation paradigms should contribute in achieving the optimal functionality. 
Optimal stimulus settings (such as amplitude, pulse frequency and duration, amplitude increase and fade time, as well as polarity, current density, and charge balance) should be determined for each particular neurostimulation case to achieve optimal control, taking into account the anatomical and functional organization of the neural network and the characteristics of conducting current in a given tissue area.
Stimulation settings should be determined in regard to safety in order to eliminate potential adverse effects on tissues. 
An adequate power supply is important to adapt the neurostimulator to daily use, providing an appropriate degree of freedom. 
The limitation associated with connection to an external battery represents an important barrier to full functional independence. 
In addition, many current neurostimulation strategies still using direct connection of the external part of the electrode with a stimulator and a power source, commonly used for trial stimulation. 
In contrast, fully implantable power supplies eliminate external constraints, thus providing mobility and reliability. 
Implantable power supplies must provide sufficient capacity to independently deliver continuous power to the neurostimulator, however, are time-limited and require regular replacement. 
Clinically used relatively low current devices, such as pacemakers or deep brain stimulation, have already reached suitable battery capacity (5-10 years) for convenient use, which, if not optimal, is generally acceptable for clinical use in most cases. 
Alternatively, implanted secondary power supplies (i.e. rechargeable batteries) provide an excellent set of advantages over the main power supplies. 
Multiple currently available on market devices also have option of recharging or using external device as a source of continuous power, which come with pros and cons, like recharging time or importance of having external part close to the implant site. 
Considerations about the optimal power supply for neuroimplants are primarily limited to the amplitude characteristics necessary for stimulation and energy efficiency of the stimulator, the characteristics of the new miniaturized implanted primary or secondary batteries, as well as the desired life and working cycle of the battery, in combination with individual settings. 
Systems for transmitting signals must be included in the form of real-time wireless communication both between components of the device (for example, a control system stimulator), or other devices (for example, BCI-to-neuroprosthesis, neuroprosthesis to feedback systems, etc.), as well as external communication for applications such as device maintenance, control modification and adaptation through training algorithms, or using telemedicine. 
Therefore, signaling requires the capability of real-time high-speed wireless data transmission. 
Telemetry can be achieved through a variety of signal transmission modalities, such as radio frequency, Bluetooth, GSM, Wi-Fi, etc. 
Reliable, safe, and trouble-free transmission of a radio signal between devices makes proper stability of transmission and data security a basic condition. 
Signal transmission must be accurate and reliable without interruption and protected from internal or external interference and from signals that are not connected to the device.

Surgical approaches are critical for the rapid and safe introduction of neurostimulators and their translation into clinical practice. 
The main approaches require direct access or even penetration into the tissue of the nervous system, and are accordingly associated with risk, compared to peripheral stimulation methods. 
Deep brain stimulation, in particular, requires the destruction of the blood-brain barrier, disrupting the hard brain, as well as invasive penetration into the central nervous system, and thus entails the risk of damage to neuronal structures. 
Therefore, minimally invasive methods of delivering maximally atraumatic stimulating electrodes are required to provide safety for implantation  of such neuronal interfaces. 
To be effective and safe, surgical methods must be improved to successfully implant electrodes with a reasonable margin of safety and with minimal risk of damage. 
Most importantly, the level of invasiveness should be justified by superior functional improvement from stimulation. 
The safety and strength of implants is critical for the long-term operation of neuro-prostheses in chronic implantation. 
Importantly, most spinal cord injuries occur in young adults, in 50\% of cases, before the age of 30. 
Therefore, given the lifespan continuing to improve in people with SCI and approaching the rest of the population, these patients will be able to potentially benefit from neuroprostheses for 50 years or more. 
To provide stable improvement and comparative   functional independence for patients, implantable devices must have independent functionality with minimal maintenance required. 
The requirements of strength, reliability, and functionality over a long period of time, the reliability of the hardware, as well as biological compatibility, are extremely important issues. 
Implantable devices must be reliable, providing them with chemical and electrical stability, as well as resistance to the potentially corrosive effects of extracellular fluid and the effects of the immune system. 
The devices must have a minimum weight, volume, and provide adequate isolation from extracellular fluid to protect the sensitive circuit, as well as tissues must be reliably protected  from potential toxic reactions caused by the neuroimplant. 
In addition to chemical insulation, electrical insulation is also important to ensure high stimulation   efficiency without current leakage. 
The individual components of the system, such as the stimulator, terminals, connectors, and electrodes, must be sufficiently flexible to provide appropriate mechanical rigidity (for brittle parts of the stimulator) and flexibility (for electrodes) to achieve optimal performance, duration of use and resist wear. 
Finally, on the basis of practical considerations, the devices should be fully implanted with all components and should include a control system, a stimulator, and an energy source.

In conclusion, current neuroimplants and neurostimulators provide a wide range of advantages over devices with external components and are therefore considered more reliable for chronic use~\citep{Peckham_2005}. 
Fully implantable neuroprostheses provide (1)~systemic resistance and superior reliability, (2)~implantation eliminates the need to control inconvenient external system components, (3)~the implantable device is optimal for comfort, cosmetic effect, and mobility, (4)~the implantable device reduces the risk of malfunctioning, risk of mechanical deformation, and risk of systemic infection, (5)~the cumulative benefits of implanted devices tend to simplify their use in future.

\section{Conclusion}\label{sec:conclusion}

In this work we proposed new concept of ``neuropunk revolution'' as the paradigm shift and view on the real-time neurosimulations and their integration into the nervous and other biological systems. 
We see promising perspectives as novel scientific approach as well as software and hardware implementations including novel electronic devices (memristors). 
The use of a real-time neurosimulation to replicate one or several functions of the mammalian nervous system provides freedom to integrate it back into biological nervous system using modern BCI technologies or to use as the part of novel robotic system including neurorehabilitation exoskeletons. 
One of the brightest examples is the closed loop system to compensate the loss of motor activity of paraplegic patients with spinal cord injury. 
In this case we could integrate the information from digital pressure sensors and goniometers mounted on legs of patients into the real-time neurosimulation of the spinal cord segment where the compensating neuronal activity could be generated and later via muscles or neurostimulation integrated into the biological nervous system that in its turn could facilitate the reconstruction of the locomotion spinal cord circuitry.

\section{Acknowledgment}

The authors would like to thank the B-Rain Labs LLC company for supporting the work in neurosimulations and personally Andrew Kuptsov. Prof. Vallverd\'u research is supported by an ICREA Acad\'emia grant. 
Several authors from Kazan Federal University have been supported by the Kazan Federal University Strategic Academic Leadership Program (``PRIORITY-2030'').
Research devoted to hardware implementation of memristive high-dimensional neurons (subsection 3.1.1) is supported by the Russian Science Foundation (grant No. 21-11-00280, https://rscf.ru/project/21-11-00280/). Development of memristive neuromorphic and neurohybrid systems (section 3.2) is supported by the Government of the Russian Federation (Agr. No. 074 02 2018 330 (2)). All authors from Lobachevsky University have been supported by the Lobachevsky University Strategic Academic Leadership Program (``PRIORITY-2030''). All authors from Baltic Federal University have
been supported by the Immanuil Kant Baltic Federal University Strategic Academic Leadership Program (“PRIORITY-2030”).

%\bibliography{references, bibliography}% common bib file


\end{document}